\documentclass[11pt]{article}

\usepackage{amssymb, amsmath,amsthm, mathrsfs, natbib}
\usepackage{ulem}
\usepackage{latexsym}
\usepackage{epsfig}
\usepackage{booktabs}
\usepackage{graphicx}
\usepackage{xcolor}
\usepackage{amsmath,amsfonts}
\usepackage{float}
\usepackage{natbib}
\usepackage{subcaption,authblk} 
\usepackage{multirow}
\usepackage[colorlinks=true, citecolor=blue]{hyperref}

\usepackage[margin=1in]{geometry}
\usepackage{setspace}
\newtheorem{Proposition}{Proposition}

\newtheorem{Theorem}{Theorem}

\newtheorem{Remark}{Remark}

\newcommand{\be}{\begin{equation}}
\newcommand{\ee}{\end{equation}}
\newcommand{\beqa}{\begin{eqnarray*}}
\newcommand{\eeqa}{\end{eqnarray*}}
\newcommand{\beqn}{\begin{eqnarray}}
\newcommand{\eeqn}{\end{eqnarray}}
\newcommand{\baa}{\begin{array}}
\newcommand{\eaa}{\end{array}}
\newcommand{\bcc}{\begin{center}}
\newcommand{\ecc}{\end{center}}
\newcommand{\btab}{\begin{tabular}}
\newcommand{\etab}{\end{tabular}}

\newcommand{\mb}{\makebox}

\newcommand{\lb}{\label}

\newcommand{\msC}{\mathscr{C}}

\newcommand{\mcC}{\mathbb{C}}

\newcommand{\mcH}{\mathcal{H}}

\newcommand{\mcL}{\mathcal{L}}
\newcommand{\mcM}{\mathcal{M}}

\newcommand{\mcV}{\mathcal{V}}

\newcommand{\diag}{{\rm diag}}

\newcommand{\argmin}{\mb{\rm argmin}}

\newcommand{\btheta}{\boldsymbol{\theta}}
\newcommand{\bTheta}{\boldsymbol{\Theta}}

\newcommand{\blambda}{\boldsymbol{\lambda}}
\newcommand{\beeta}{\boldsymbol{\eta}}
\newcommand{\bbeta}{\boldsymbol{\beta}}

\newcommand{\bpsi}{\boldsymbol{\psi}}
\newcommand{\bXi}{\boldsymbol{\Xi}}
\newcommand{\bxi}{\boldsymbol{\xi}}

\newcommand{\bSigma}{\boldsymbol{\Sigma}}
\newcommand{\bPsi}{\boldsymbol{\Psi}}

\newcommand{\bmu}{\boldsymbol{\mu}}

\newcommand{\bA}{\boldsymbol{A}}
\newcommand{\bB}{\boldsymbol{B}}
\newcommand{\ba}{\boldsymbol{a}}

\newcommand{\bC}{\boldsymbol{C}}
\newcommand{\bc}{\boldsymbol{c}}
\newcommand{\bD}{\boldsymbol{D}}

\newcommand{\bv}{\boldsymbol{v}}

\newcommand{\bH}{\boldsymbol{H}}

\newcommand{\bL}{\boldsymbol{L}}
\newcommand{\bI}{\boldsymbol{I}}

\newcommand{\bt}{\boldsymbol{t}}

\newcommand{\bF}{\boldsymbol{F}}

\newcommand{\bP}{\boldsymbol{P}}

\newcommand{\bW}{\boldsymbol{W}}

\newcommand{\bx}{\boldsymbol{x}}
\newcommand{\bX}{\boldsymbol{X}}

\newcommand{\bY}{\boldsymbol{Y}}

\newcommand{\bz}{\boldsymbol{z}}

\newcommand{\bzero}{\boldsymbol{0}}
\newcommand{\bone}{\boldsymbol{1}}

\newcommand{\EE}{{\mathbb E}}

\newcommand{\RR}{{\mathbb R}}

\title{ Fully Bayesian Benchmarking for Small Area Models with Count Data}
\author[1]{Soumadeep Bhowmick}
\author[1,3]{Anindya Roy}
\author[2,3]{Gauri S. Datta}
\author[3]{Kyle M. Irimata}
\author[3]{Jerry J. Maples}

\affil[1]{University of Maryland Baltimore County, Baltimore, MD, USA}
\affil[2]{University of Georgia, Athens, GA USA}
\affil[3]{US Census Bureau, Washington, D. C. USA\footnote{Disclaimer: This report is released to inform interested parties of research and to encourage discussion. The views expressed on statistical issues are those of the authors and not those of the U.S. Census Bureau.}}

\usepackage{titling}
\pretitle{\begin{center}\LARGE\bfseries}
\posttitle{\par\end{center}\vspace{-0.6em}}

\preauthor{\begin{center}\large}
\postauthor{\par\end{center}\vspace{-0.2em}}
\date{}
\begin{document}

\maketitle
\vspace{-0.6in}
\begin{abstract}
Small area estimates of counts are often required to conform exactly to reliable aggregates at a higher geographic level. Often the area-level model is specified on a transformed (nonlinear) scale such as a log scale.  Specifying the small area model on the log-scale is common practice for positive responses. A  prominent example is the production of the Small Area Income and Poverty Estimates (SAIPE)  by the US Census Bureau. When the model is specified in a transformed scale, the benchmarking constraints, which are linear in the area means, become nonlinear in terms of the model parameters. We develop a new class of priors for Bayesian small area models under strict external benchmarking constraints when the small area model is specified in a transformed scale, and the model parameters belong to a Riemannian manifold due to the nonlinear benchmarking constraints. 
The prior is fully supported on the constrained parameter set, the manifold defined by the benchmarking constraints and thereby provides a proper constrained posterior that is also supported on the restricted set. This facilitates full Bayesian analysis of the problem without the requirement of post-hoc modification to the posterior to enforce the benchmarking constraints.  \\
{\bf Keywords:} External benchmarking; Fay-Herriot model; Interior Point Method; Optimization; Smooth Manifold. 
\end{abstract}

\newpage

\section{Introduction}

Small area estimation (SAE) is concerned with obtaining reliable estimates of population characteristics for geographic or demographic domains when  direct survey estimates for those  areas are subject to substantial sampling variability  due to limited sample sizes  associated with the direct surveys. This is  a common problem in  official statistics,
public health, social science, and economic applications, where estimates are required for counties, municipalities, demographic subgroups, or other domains that are substantially smaller than the populations represented in the parent
survey. A central principle of model-based small area estimation is therefore to borrow strength across areas through auxiliary information and an
appropriate statistical model. 

The Fay-Herriot (FH) model introduced by
\cite{FayHerriot1979} remains one of the principal area-level models that is used to generate model-based estimates. Comprehensive treatments of small area estimation and the development of the FH methodology are given by \cite{RaoMolina2015} and \cite{Pfeffermann2013}. Often for social and economic variables,  the behavior of the measurements is  multiplicative. When these measurement are done  using sample surveys, the survey errors also become multiplicative. In such cases, to use standard FH-type model-based estimation, it is common to  log-transform the measurements to convert to an additive scale. This is a typical scenario in small area estimation exercises  with count data. Counts are nonnegative, often highly variable across areas, and can exhibit substantial right skewness, making the conventional Gaussian FH model potentially inappropriate in the original scale. A natural approach in applications involving positive count responses is therefore to model the log transformation of the response. A transformed FH model retains the attractive features of the FH framework while allowing multiplicative rather than additive variation on the original scale. Such models provide a convenient framework for borrowing strength across areas when the underlying area-specific quantities are positive. The statistical consequences of the transformation, including retransformation bias and mean squared error estimation, have been studied in the small area literature; see, for example, \cite{MaitiPrasad2006} and \cite{ChandraSud2017}.

An important feature of many small area applications is that estimates for individual areas are required to be consistent with reliable information available at a higher level of aggregation. For example, an administrative data source, a census, or a large-scale survey may provide an externally determined estimate of a total or weighted average over a collection of small areas. In such settings it is desirable that the final small area estimates satisfy an externally supplied benchmark. Typical  benchmarks have the form
\[\sum_{i=1}^m w_{j,i} \mu_i = t_j, \;\; j=1, \ldots,q,\]
where $w_{j,i}$ are known aggregation weights corresponding to the $j$th  benchmarking constraint, $t_j$ is the benchmark obtained independently of the small area survey pertaining to the $j$th constraint and $\mu_i$ is the  mean associated with  area $i$ for the response of interest. 

The concept of benchmarking has a long history  in the literature on time series; see \cite{Denton1971, Cholette1984}. The classical paper by \cite{hillmer1987benchmarking} followed up by \cite{TrabelsiHillmer1990,  durbin1997benchmarking}  analyzed benchmarking in the context of model-based estimation. In the small area literature, early papers by  \cite{PfeffermannBarnard1991} and \cite{wang2008small}  studied constrained estimation under benchmarking.   The general objective of benchmarking in the context of small area models  is to modify model-based small area estimators so that they agree with a more reliable aggregate estimate while retaining, as far as possible, the efficiency and shrinkage properties of the underlying model. Classical
work has considered constrained estimation under the Fay-Herriot model, while subsequent developments have studied optimal benchmarked estimators, empirical Bayes benchmarking, multiple benchmark constraints, and mean
squared error estimation. For example, \cite{WangFullerQu2008} studied small area estimation under an aggregate restriction, while \cite{BellDattaGhosh2013} developed a general framework
for benchmarking small area estimators under linear constraints. \cite{SteortsGhosh2013} investigated the mean squared error of benchmarked empirical Bayes estimators.  From a Bayesian perspective \cite{datta2011bayesian} developed a general class of benchmarked estimators by minimizing posterior expected loss subject to the benchmark constraints which was further extended to accommodate multilevel benchmarking in \cite{GhoshSteorts2013}. However, these approaches produce point estimates with the benchmarking property but not a full posterior distribution that is supported on the constrained parameter space, hence uncertainty quantification frameworks must be developed separately. 

Fully Bayesian benchmarking has gained attention as an approach to incorporating benchmarking constraints into Bayesian inference. It seeks a posterior distribution such that it is supported in the constraint parameter space, which means that any posterior draw from this joint distribution will satisfy the benchmarking constraints. \cite{zhang2020fully,okonek2024computationally} enforce these constraints through linear adjustments within the Metropolis–Hastings step during posterior sampling. Another approach by \cite{sugasawa2024bayesian} uses entropic tilting to construct an approximate posterior distribution whose expectation satisfies the benchmarking constraints by reweighting unconstrained posterior samples. This approach imposes constraints on the posterior expectation rather than on the entire posterior support. Alternatively, \cite{bhowmick2026benchmarkingpriorsfullybayesian} specify a class of priors supported in the constrained parameter space, ensuring that the resulting posterior satisfies the benchmarking constraints.

The existing benchmarking literature, however, has been developed predominantly for constraints that are linear in the small area parameters being estimated.  The paper by \cite{durbin1997benchmarking}  studied benchmarking under both additive and multiplicative models. However,  the methodology of benchmarking small area model-based estimators seems to be more developed for the Gaussian case where the benchmarks on the model parameters are linear constraints.  The need for handling nonlinear constraints becomes unavoidable when the area-level model is specified on a transformed scale  and $\mu_i$, the mean response on the original scale becomes a nonlinear function, $\mu_i(\theta_i),$ of the parameters $\theta_i$ in the transformed scale.  We consider a single benchmarking constraint for illustration, but the same idea holds for multiple constraints. In the present setting, the small-area model for direct estimates is linear in the logarithmic parameter $\theta_i$, but the external benchmark is
defined on the original scale through
\[\sum_{i=1}^m w_i \exp(\theta_i) = t.\]
Thus, although the  small area model is a Gaussian linear mixed model, the scientifically meaningful benchmark defines a nonlinear equality constraint on the model parameters. This distinction is fundamental. A constraint imposed on latent parameters $\theta_i$, $ \sum_{i=1}^m w_i\theta_i = c,$ %[\GSD{Please check the constraint; it is not on log}] %updated the statement(SB)
is generally not sufficient to achieve the desired benchmark. In particular,
\[
    \exp\left\{\sum_{i=1}^m w_i\theta_i\right\}
    \neq
    \sum_{i=1}^m w_i\exp(\theta_i).
\]
Therefore, a procedure that applies a conventional linear benchmarking adjustment on the logarithmic scale may produce estimates that satisfy a constraint mathematically but fail to satisfy the externally specified aggregate requirement on the original scale.

The present paper develops a framework for externally benchmarked small area estimation under log-scale Fay-Herriot model for area-level count data. The central feature of the proposed formulation is that the benchmark is imposed  directly on the original-scale area-specific means rather than on the transformed Fay-Herriot parameters. This produces a nonlinear benchmarking constraint in the underlying Gaussian Fay-Herriot parameters. The resulting framework provides a natural synthesis of transformed small area modeling and external benchmarking.

Our approach is motivated by three considerations. First, the logarithmic Fay-Herriot model provides a parsimonious mechanism for modeling heterogeneous
positive area-level counts while borrowing strength across areas. Second, external benchmark information is naturally available on the original count or population scale rather than on the logarithmic scale. Third, requiring agreement with an externally supplied aggregate can be viewed as a coherence requirement between model-based small area estimates and independently produced aggregate information. The proposed methodology therefore seeks to retain the statistical advantages of Fay-Herriot shrinkage while respecting an externally determined aggregate constraint on the scientifically meaningful scale.

The proposed framework also differs conceptually from conventional post-hoc benchmarking. In the standard approach, an unconstrained collection of small area predictors is first obtained and subsequently adjusted to satisfy a linear aggregate constraint. In the present setting, the benchmark is itself a nonlinear function of the model parameters. Treating the nonlinear constraint directly incorporates the external information into the benchmarking problem rather than treating it solely as an after-the-fact calibration condition. This distinction has consequences for both point prediction and uncertainty quantification because the benchmark couples the small area parameters across domains.

The methodological development contributes to the literature on benchmarking for small area models in several related directions. First, it extends transformed Fay-Herriot modeling for positive and count-valued area-level responses by explicitly distinguishing the logarithmic modeling scale from the original scale on which the target is defined. Second, it extends classical benchmarking methodology from linear constraints on small area parameters to nonlinear constraints induced by the inverse transformation of a Gaussian area-level model. Third, it provides a framework for studying the statistical consequences of imposing externally supplied aggregate information, including its effects on shrinkage, prediction error, and uncertainty quantification. Finally,   we present  theoretical results for benchmarking prior construction under a general setting where the response for the Fay-Herriot model is a smooth transformation of the direct area-level estimates. While this paper focuses on the specific application of SAE for poverty count from SAPIE, which uses a lognormal transformation of the direct estimates,  the proposed benchmarking methodology is equally applicable for broad  other range restricted responses such as area-level proportions which may use a logit or probit transformation.

\subsection{Motivating Application: SAIPE}

A particularly important application of small area estimation arises in the estimation of county-level poverty in the United States. The U.S. Census Bureau's SAIPE program produces annual model-based estimates of income and poverty for every county in the United States, as well as for states and school districts. The program is designed specifically to provide reliable single-year estimates for small geographic areas for which direct survey estimates may have substantial sampling variability. SAIPE estimates are used not only for statistical description of local economic conditions but also in the administration of federal programs and the allocation of federal funds to local jurisdictions and school districts. The importance of accurate and stable county-level poverty estimates is therefore both statistical and substantive. In particular, relatively small changes in estimated poverty counts or rates can have consequences for the assessment of local economic conditions and for resource allocation across geographic areas.

The SAIPE program illustrates particularly well the fundamental motivation for small area modeling. County-level poverty information is available from the American Community Survey (ACS), but direct estimates for smaller counties can be substantially less precise than estimates for larger geographic areas. The 5-year ACS direct estimates are released for all counties; however, the 1-year direct estimates are only publicly released for counties with a population of at least $65,000.$ The Census Bureau consequently combines ACS information with administrative and other auxiliary data, including information derived from federal income tax returns, SNAP participation, decennial census data, postcensal population estimates, and other economic indicators. The resulting SAIPE estimates are model-based rather than direct enumerations or direct survey estimates. The Census Bureau notes that the resulting estimates generally have smaller sampling error than corresponding direct survey estimates and are particularly useful for counties with smaller populations. These features make county-level poverty estimation a natural setting for area-level models that borrow strength across counties. The SAIPE methodology underscores the importance of considering coherence of the small area estimate with reliable information from broader geographic levels along with accuracy of the estimates,  particularly for downstream  decision making. 

The SAIPE program produces estimates for children under 18 years of age pov18, but the estimates are further broken into two groups,  estimates for children less than  5 years old, pov5, and for children between 5 and 18 years old, pov517. Thus, at the county-level one has the opportunity to benchmark these two groups to the overall SAIPE estimates for the under 18 age group, i.e., pov18 = pov5 + pov517 for each county level estimates. While not a part of the US Census Bureau  benchmarking procedure for the SAIPE production process, such further nested benchmarking can improve the finer scale estimates, and thus a procedure that can handle the nested hierarchical structure of benchmarks is desirable.

\section{Small Area Model}
We  develop the small area model and methodology, keeping SAIPE as our motivating example. Hence we prescribe the model for the direct count estimates in the log-scale; the methodology, however, can handle more general transformations.  Suppose $Y_i$ denotes the observed count for small area $i$, $i=1,\dots, m$. We assume a log–linear model for the counts,
\be
\log Y_i = \theta_i + \epsilon_i, \qquad \epsilon_i \sim N(0,\sigma_{i,e}^2),
\lb{eq:FayHerriot}
\ee
where $\theta_i$ represents the latent small-area effect. In practice, $\theta_i$ may depend on auxiliary covariates through a regression structure. In the classical Fay-Herriot model (\cite{FayHerriot1979}), the linking of the area means on the area-level covariates is done using a linear structure  \be
\theta_i = \bx_i^{T}\bbeta + u_i, \qquad u_i \sim N(0, \sigma_u^2)
\lb{eq:Fay-Herriot}
\ee
where $\bx_i = (x_{i,1}, \ldots, x_{i,p})^\top$ is a $p\times 1$ vector of known predictors, $\bbeta$ is a vector of regression coefficients, and $u_i$ are random area effects.  
%\GSD{Did we introduce the distribution of $u_i$?} 
The hierarchical model specification is usually completed by specifying priors on the coefficients $\bbeta$ and the random effects variance $\sigma_u^2,$ 
\[  \pi(\bbeta) \propto 1, \qquad  \pi(\sigma_u^2) = \mathrm{IG}(a_0, b_0),\]
where $\mathrm{IG}(a_0, b_0)$ is the inverse-gamma density  with the kernel $\exp(-b_0/\sigma_u^2)(\sigma_u^2)^{(-a_0 -1)}$.

Let $\bY = (Y_1,\dots,Y_m)^T$ denote the vector of small-area counts and let $\bA$ be a known $q\times m$ aggregation matrix defining higher-level totals. For example, $\bA$ may map county-level counts to state-level totals. Let $\bt \in \RR^q$ denote the known benchmark totals obtained from reliable external sources.
The benchmarking requirement is that the expected aggregated counts implied by the model must agree with the benchmark totals. Formally, we impose the constraint
\[
\bA\EE(\bY \mid \btheta) = \bt.
\]
Because the model operates on the log scale, using the  moment formula for the log-normal distribution the conditional expectation of $Y_i$ given $\theta_i$ is
$\EE(Y_i \mid \theta_i) = \exp(\theta_i + \frac{1}{2}\sigma_{i,e}^2).$
Therefore, the benchmarking constraint becomes $
\bA\bD\exp({\btheta}) = \bt, $ where the exponential is componentwise, i.e., $\exp{(\btheta)} = (\exp(\theta_1),\dots,\exp(\theta_m))^T$ and $\bD = \diag(e^{\frac{1}{2}\sigma_{1,e}^2}, \ldots, e^{\frac{1}{2}\sigma_{m,e}^2}).$ Writing the $q \times m$ matrix of  constraints $\bA\bD$ as  $\bW = \bA\bD,$ the parameter  space under the hard external benchmarking constraints becomes 
\be
 \mcM_m = \{\btheta \in \RR^m: \bW\exp{(\btheta)} = \bt\} .
\label{eq:benchmark}
\ee
While the unconstrained prior specification of the  standard Fay-Herriot model still provides Bayesian inference for the area mean vector $\EE(Y_i \mid \theta_i),$ Bayesian inference becomes intractable under the nonlinear benchmarking constraints. Often the procedures use post-hoc modification to the posterior of $\btheta$ to enforce the benchmarking constraints, thereby moving away from a full Bayesian analysis on the quantities of interest. 

Rather than enforcing benchmarking through post-hoc adjustments, we propose to incorporate the benchmarking constraint directly within the Bayesian model. Our objective will be to  construct priors  that are fully supported on the constrained parameter space $\mcM_m$. Such  prior  would lead  to automatically benchmarked  posterior giving estimates and uncertainty quantification both under exact benchmarking constraints. 

Under this fully Bayesian approach, posterior inference simultaneously respects the hierarchical model structure and the external benchmarking information. This allows information from both sources to be propagated through the posterior distribution of the small-area counts.
Embedding benchmarking constraints within the Bayesian hierarchy offers several advantages. First, it guarantees coherence between small-area estimates and aggregate benchmarks at every iteration of the posterior sampler. Second, uncertainty in the latent parameters is modulated via the benchmarking constraint rather than ignored through deterministic adjustments. Bnechmarking priors for the linear Gaussian  Fay-Herriot model were investigated in \cite{bhowmick2026benchmarkingpriorsfullybayesian,fúquenepatiño2026regularizedsmallareaestimation}.

\section{Exact Benchmarking Under General Nonlinear Transformation}
\label{sec:general-hard-benchmark}

In many applications, the  area level   direct estimates  are  range  restricted, e.g.,  for count response the range is usually $(0, \infty)$ (excluding  the zero count cases), or $(0, 1)$ when the response is a proportion.  In such cases, one option is to use a generalized linear model formulation to  borrow strength by  linking the auxiliary information to the  mean of the response  under exponential family  models. However, given the extensive use of the Gaussian  Fay-Herriot  model, a common option remains to be use the Gaussian model under a transformed  response. For example, in our  motivating application for positive counts,  the Gaussian Fay-Herriot model  can be specified in terms of the logarithm of the counts,  resulting in a log-normal  distribution for the counts. 
Specifically,  let ${\tilde{Y}}_i$ be the direct estimate for area $i$ and let $Y_i = T({\tilde{Y}}_i)$ be a  component wise  transformation which  removes any range restriction from the  transformed response $Y.$

In such transformed model, the external benchmark may be expressed in terms of the mean of the original  direct  estimates, $\mu_i(\theta_i) =  E({\tilde{Y}}_i | \btheta)$  where the  $\mu_i$  functions  may depend on the  entries of the  sampling error  covariance matrix $\bSigma_{e}.$  This necessitates consideration of benchmark constraints of the form $\bW^\top \bmu(\btheta)= \bt,\; \bt \in \RR^q,$ where $\btheta=(\theta_1,\ldots,\theta_m)^\top$ and $\bmu(\btheta) = (\mu_1(\theta_1), \ldots,   \mu_m(\theta_m))^\top.$
Here $\bW \in  \RR^{m\times q}$ is assumed to have full column rank.  Define the benchmark functional
\be
\bF(\theta) = \bW^\top \bmu(\btheta) - \bt : \RR^m \to \RR^q.
\label{eq:F-general}
\ee
The hard  external-benchmark parameter space is therefore
\be
\mcM_m = \bF^{-1}\{\bzero\} = \left\{ \btheta\in\RR^m:
\bF(\btheta) = \bzero \right\}
= \left\{ \btheta: \bW^\top \bmu(\btheta) = \bt
\right\}.
\label{eq:general-manifold}
\ee
Then, the $q\times m$ Jacobian matrix of $\bF$ is 
\be
D \bF(\btheta) = \bW^\top J_{\mu}(\btheta),
\label{eq:DF-general}
\ee
where $J_{\mu}(\btheta) = D \bmu(\btheta) = \diag \left\{ \mu_1'(\theta_1),\ldots,\mu_m'(\theta_m)\right\}$ is the $m\times m$  Jacobian matrix  for $\bmu : \RR^m \to \RR^m.$ 
We will throughout make the global assumptions 
\begin{enumerate}

\item[(G1)] 
The coordinate maps $\mu_i$  satisfy $\frac{\partial \mu_i(\theta)}{\partial \theta_i} > 0$  for all $\theta \in \mcM_m$  for each $i = 1, \ldots,m.$ 

\item[(G2)] 
The map $\bmu:  \RR^m \to \RR^m$ is $C^1$, and
$\operatorname{rank} \left\{\bW^\top J_{\bmu}(\btheta) \right\} = q$ for every $\btheta\in\mcM_m.$
\end{enumerate}
To obtain the  unrestricted  coordinates on the manifold to which  priors can be specified, we  
assume  $\theta^\circ \in \mcM_m$ is an  interior point with $\bW^\top \bmu(\btheta^\circ) = \bt,$ and consider the representation
\be
\btheta(\beeta) = \btheta^\circ + \bB\beeta -
\bC\blambda(\beeta),
\qquad
\beeta\in\RR^{m-q},
\label{eq:chart}
\ee
where $\bC \in  \RR^{m\times q}$ and $\bB \in \RR^{m\times(m-q)}$ is a matrix  whose columns   form an  orthonormal basis of $\ker(\bW^\top)$, i.e.,  $\bW^\top\bB = \bzero$ and  $\bB^\top\bB = \bI_{m-q}.$ We will assume $\msC(\bC) = \msC(\bW)$ and hence
\begin{enumerate}
    \item[(P1)]  $[\bB : \bC]$ is  nonsingular. 
\end{enumerate} 
The  tuning parameter  $\blambda(\beeta)\in\RR^q$ is chosen so that
\be
\bW^\top \bmu( \btheta^\circ + \bB\beeta - \bC\blambda(\beeta))  = \bt.
\label{eq:lambda}
\ee
Thus $\beeta$ represents the $m-q$ free directions, while $\blambda(\beeta)$ adjusts the remaining $q$ directions to satisfy the hard
benchmark.

% Condition \eqref{eq:nonzero_det} is the general analogue of the positive definiteness condition that arises in the linear  case  of Gaussian  models. It guarantees that movement in the $\bC$ directions changes the benchmark constraints in $q$ locally independent ways, so that the $q$ equations can indeed be solved for the $q$ coordinates of $\blambda$.

The following four results establish progressively that \eqref{eq:chart} indeed yields a regular local coordinate system for $\mcM_m$. Finally we will establish a global coordinate system under additional assumptions on the maps. 

\begin{Proposition}
\lb{prop:regularity}
Assume the regularity  assumption (G2). Then in a neighborhood of $\btheta^\circ,$ $\mcM_m$ as defined in \eqref{eq:general-manifold} is an embedded $C^1$ submanifold of $\RR^m$ of dimension $m-q$.  Moreover, for every
$\btheta\in\mcM_m$ in that neighborhood,
\[ T_{\btheta}\mcM_m = \ker \left\{
D \bF(\btheta) \right\} =
\ker \left\{ \bW^\top J_{\bmu}(\btheta) \right\}.
\]
\end{Proposition}

The original parameter $\theta$ contains $m$ coordinates, while the $q$ independent hard
benchmark constraints remove $q$ local degrees of freedom.  Consequently, an exact reduced parameterization should involve precisely $m-q$ unrestricted coordinates. Next we show that a solution $\blambda(\beeta)$ to \eqref{eq:lambda}  exists  locally and smoothly. 

\begin{Proposition}
\lb{prop:IFT-lambda}
Define
\be
\Phi(\beeta,\blambda) = \bW^\top \bmu( \btheta^\circ + \bB\beeta - \bC\blambda) - \bt.
\label{eq:Phi}
\ee
Since $\btheta^{\circ}\in\mcM_m,$ we have
$\Phi(\bzero,\bzero)=\bzero.$ Assume  (G1).  
Then there exist neighborhoods $U \subset \RR^{m-q}$ of $\bzero \in  \RR^{m-q}$ and
$V \subset \RR^q$ of $\bzero \in \RR^q,$
and a unique continuously differentiable function
$\blambda:U\to V$ such that $\blambda(\bzero)=\bzero$ and
$\Phi \left\{ \beeta, \blambda(\beeta) \right\} =
\bzero,$ for every $\beeta\in U.$ Consequently,
\[
\btheta(\beeta) = \btheta^{\circ} +
\bB\beeta  - \bC\blambda(\beeta)
\]
is a well-defined $C^1$ map from $U$ into $\mcM_m$.
\end{Proposition}
Thus the implicit function theorem establishes a smooth $(m-q)$-parameter map into the benchmark manifold. However, to show a full local coordinate we need to show the  mapping spans the entire tangent plane at $\btheta(\beeta).$

\begin{Proposition}
\lb{prop:H-tangent}
Under the assumptions of Propositions
\ref{prop:regularity} and \ref{prop:IFT-lambda}, define
$\bH(\beeta) =  \frac{\partial \btheta(\beeta)}{\partial\beeta^\top}.$
Then 
\[ \msC\{H(\beeta)\} = T_{\btheta(\beeta)}\mcM_m.\]
\end{Proposition}
The proposition shows that $\btheta$ is a local  diffeomorphism  at each point $\beeta$. We can now show that the reduced parameter  $\beeta$, i.e., the inverse  map $\bTheta^{-1}$ defines a regular local coordinate system.

\begin{Theorem}
\lb{thm:local-chart}
Suppose the assumptions of Propositions
\ref{prop:regularity}--\ref{prop:H-tangent} hold.  Then there exists an open neighborhood $U_0\subset\RR^{m-q}$ of $\bzero$ and an open neighborhood
$\mcV_0\subset\mcM_m$ of $\btheta^{\circ}$, where openness is relative to
$\mcM_m$, such that the map
\be
\bTheta: U_0\longrightarrow\mathcal V_0, \qquad
\bTheta(\beeta) = \btheta^{\circ} + \bB\beeta - \bC\blambda(\beeta),
\label{eq:Theta-map}
\ee
is a $C^1$ diffeomorphism. Consequently,
$\beeta \in \RR^{m-q}$ provides a regular local coordinate system for the hard-benchmark manifold near $\btheta^{\circ}$.  In particular, every $\btheta \in \mathcal V_0$ has a unique representation
\[
\btheta = \btheta^{\circ} + \bB\beeta - \bC\blambda(\beeta)
\]
for a unique $\beeta\in U_0$. Furthermore, for each $\beeta\in U_0$, the derivative $D\Theta(\beeta) = H(\beeta): \RR^{m-q} \longrightarrow
T_{\Theta(\beeta)}\mcM_m$ is a linear isomorphism.
\end{Theorem}
With additional conditions on the map we can  now establish $\beeta$ as a   global  coordinate system Assume the following conditions.
\begin{enumerate}

\item[(G3)] 
For every $\beeta\in\RR^{m-q},$
there exists unique $\blambda(\beeta)\in\RR^q$ such that
$\Phi \left\{ \beeta, \blambda(\beeta) \right\} = \bzero.$

\item[(G4)] 
At every solution,
$\det \left[ \bW^\top J_{\bmu} \left\{ \btheta^{\circ} +
\bB\beeta - \bC\blambda(\beeta) \right\} \bC \right] \neq0.$
\end{enumerate}

\begin{Theorem}
\lb{thm:global-chart}
Assume  conditions G1-G4.  Then the solution
$\beeta \mapsto \blambda(\beeta)$ is globally $C^1$.  Moreover, the map $\Theta(\beeta) =
\btheta^{\circ} + \bB\beeta - \bC\blambda(\beeta)$
is a global $C^1$ diffeomorphism $\Theta: \RR^{m-q} \to \mcM_m.$ Consequently, every $\btheta\in\mcM_m$ has a unique representation
\[\btheta = \btheta^{\circ} + \bB\beeta - \bC\blambda(\beeta), \]
and $\beeta\in\RR^{m-q}$ gives a global unrestricted
coordinate system for the entire benchmark manifold.
\end{Theorem}

\begin{Remark}
Suppose the map $\bmu(\btheta)$ is a  componentwise map, i.e., $\bmu(\btheta) = (\mu_1(\theta_1), \ldots, \mu_m(\theta_m))$  with  each  $\mu_i$ is a strictly increasing  function. This is common in the standard  Fay-Herriot  formulation under transformation for many non-Gaussian  model,  such as  under log-transformed model where $\mu_i(\theta_i) =  \exp(\theta_i).$ Then $J_{\bmu}(\btheta) = diag(\mu_i'(\theta_i))$ is positive  definite  for all $\btheta.$
If we make the canonical  choice $\bC = \bW,$ then
$\bW^\top J_{\bmu}(\btheta) \bW \succ 0.$ Hence the transversality condition (G4) is automatically satisfied.

Thus, for this particularly natural choice, the reduced representation $\btheta(\beeta) = \btheta^{\circ} + \bB\beeta - \bW\blambda(\beeta)$  provides a regular local chart whenever the benchmark transformation is
componentwise strictly increasing and sufficiently smooth.
\end{Remark} 
Henceforth,  we work  with the  canonical  choice $\bC = \bW.$

\subsection{Conditional Fay-Herriot Prior on the Nonlinear Benchmark Manifold}

Let $\pi_U(\btheta \mid \bbeta, \sigma_u^2)$
denote the ordinary unconstrained Fay-Herriot prior density. Of course, under the  benchmark restriction, any density on $\mcM_m$ is singular with respect to the Lebesgue measure on $\RR^m.$ We will define the hard-benchmark prior as the regular conditional distribution of
the ordinary Fay-Herriot prior given the benchmark functional equation, i.e. on the fiber 
\[ \bF(\theta) = \bzero, \]
or equivalently $\bW^\top \bmu(\btheta) = \bt.$
We will find the regular conditional distribution using the standard Coarea  formula;  see, for example, \cite{Federer1969} or \cite{EvansGariepy2015}.. While a  global coordinate system is not necessary for the Coarea density, we will assume all conditions of Theorem~\ref{thm:global-chart} hold and derive the regular conditional prior under that and also describe the equivalence of the regular conditional posterior to that under the reduced parameterization. The  following theorem gives the result. 

\begin{Theorem}
\lb{thm:posterior-equivalence}

Suppose all conditions from Theorem~\ref{thm:global-chart} hold and let
the map $\bmu(\btheta)$ be a  componentwise map, i.e. $\bmu(\btheta) = (\mu_1(\theta_1), \ldots, \mu_m(\theta_m))$ where for each $i$, $\mu_i(\theta_i) = \frac{d \mu_i(\theta_i)}{ d\theta_i} > 0$ for all $\theta_i.$  Consider the  global coordinate system defined by the map 
$\Theta(\beeta) =
\btheta^{\circ} + \bB\beeta - \bW\blambda(\beeta)$
where $\btheta^{\circ}$ is an interior point of satisfying $\bW^\top\bmu(\btheta^\circ) =   \bt.$
Then given  $\bbeta, \sigma_u^2$ in the Fay-Herriot model \eqref{eq:Fay-Herriot} and the hard benchmarking constraints, the regular conditional  prior, with respect to the Hausdorff measure on $\mcM_m,$  is given by  
\[
\pi_B(\btheta \mid \bbeta,\sigma_u^2) \propto
\frac{ \pi_U(\btheta \mid \bbeta,\sigma_u^2) I(\btheta \in \mcM_m)}{\left\{\det\left[\bW^\top
diag\{\mu_1'(\theta_1)^2, \ldots, \mu_m'(\theta_m)^2\}\bW\right]\right\}^{1/2}} ,
\]
where $ \pi_U(\btheta \mid \bbeta,\sigma_u^2)$ is the unconstrained prior  density given $\bbeta, \sigma_u^2.$ Conditional on $(\bbeta,\sigma_u^2)$, the posterior generated by the reduced parameterization $\beeta$ is exactly the regular conditional posterior obtained by
conditioning the unconstrained Fay-Herriot posterior on the hard benchmark:
\[\Pi_B(d\theta\mid \bY,\bbeta,\sigma_u^2) =
\Pi_U\left(d\btheta \mid \bY,\bbeta,\sigma_u^2,\,
\bW^\top \bmu(\btheta) = \bt\right).\]
\end{Theorem}
% \GSD{Many preceding equations have not been numbered. I see that all these numbers appeared in the Appendix. We need to say that.}  
The regular conditional density given by \[\pi_B(\btheta \mid \bbeta, \sigma_u^2) \propto
\frac{\pi_U(\btheta \mid \bbeta, \sigma_u^2)I(\btheta\in\mcM_m)}{
J_F(\btheta)}\]  is intrinsic to the manifold. It does not depend on the particular reduced parameterization used to describe the manifold.  The role of the chart is computational, whereas the Coarea expression characterizes the regular conditional prior geometrically. The coordinate prior given by, 
\[\pi_{B,\beeta}(\beeta \mid \bbeta, \sigma_u^2)
\propto
\pi_U(\btheta(\beeta) \mid  \bbeta, \sigma_u^2))
\frac{J_{\theta}(\beeta)}{J_F\{\btheta(\beeta)\}}
\]
is the general nonlinear analogue of the exponential-chart prior.
Choosing any other density directly on $\eta$ would in general define a different prior on the benchmark manifold.  The ratio of the chart Jacobian to the Coarea Jacobian helps preserve the  equivalence with the regular conditional Fay-Herriot prior.

\subsection{Parameterization Using  Transformed Quantity: Exponential Case}

While the  regular conditional prior identifies the target class of priors that will maintain the benchmark, the normalizing  constant is extremely difficult to obtain even for simple manifolds; thereby losing some of the attractive  features of Bayesian formulation of the regular Fay-Herriot models where partial conjugacy  can be used to speed up posterior sampling. 
An attractive alternative is to construct the prior directly on the constrained manifold through a suitable transformation and parameterization. When each component transformation is invertible, the nonlinear benchmarking constraint can be converted into a linear equality constraint in the transformed parameter space, substantially simplifying the geometry of the constrained parameter space. This approach provides a principled way to generate parameter values that satisfy the benchmarking constraint exactly, provided  the range restrictions induced by the transformations can be handled. By specifying a prior on the unconstrained coordinates and mapping it onto the manifold, one obtains a properly normalized probability distribution supported entirely on the constrained parameter space, facilitating posterior computation while preserving the benchmark by construction. 

For each $i,$   define  $\xi_i=\mu_i(\theta_i),$ and let $\bxi = (\xi_1, \ldots, \xi_m)^\top.$
The nonlinear benchmark in $\btheta$ becomes the linear constraint in $\bxi$ with the manifold  becoming 
\be
\mcM_{\xi,m} = \{ \bxi \in \RR^m: \bW^\top\bxi = \bt, \; \bmu^{-1}(\bxi) \in \mcM_m\}
\label{eq:linear-xi-benchmark}
\ee
If we choose an interior point in $\mcM_{\xi,m}$ as $\bxi^\circ$ satisfying $\bW^\top\bxi^\circ = \bt.$ One may then construct a chart in $\bxi$-space of the form
\be
\bxi(\beeta) = \bPsi\left(\bB\beeta -\bW\blambda(\beeta); \bxi^\circ\right),
\label{eq:general-Psi-chart}
\ee
where $\bPsi_m$ is chosen, i.e. $\blambda(\beeta)$ is solved to keep $\bxi(\beeta)$ inside the admissible range, $\bmu^{-1}(\bxi(\beeta)) \in \mcM_m $ where the inverse is componentwise. 
The original parameter is recovered through
$\btheta(\beeta) = \bmu^{-1}\{\bxi(\beeta)\}.$

In our specific motivating application involving the Fay-Herriot model for the count data in the log-scale, the  parameterization corresponds to
$\mu_i(\theta_i)  = e^{\theta_i}$ and  
$\bPsi(\bz;\bxi^\circ) = \bxi^\circ\odot e^{\bz}.
$ Thus
\[
\bxi(\beeta) = \bxi^\circ \odot \exp \left\{
\bB\beeta - \bW\blambda(\beeta) \right\}.
\]
The exponential function is  particularly attractive because it simultaneously guaranties positivity, global invertibility, smoothness,
and a particularly simple derivative structure.
Moreover, as we will see,  it also guaranties a global chart by satisfying the solvability condition (G3) in Theorem~\ref{thm:global-chart}.

\section{Benchmarking Prior for Exponential Benchmarking Constraints}
The Fay-Herriot model for the count data is prescribed in the log scale. To this end, we focus on $\exp{(\btheta)}$  for parameterization. 
As noted earlier, in this situation the benchmarking constraints can be reduced to of the form $\bW e^{\btheta} = \bt.$
Let $\bxi = \exp{(\btheta)}.$
Then  $\bxi \in \RR^m$ belongs to  the constrained set
\be 
\mcM_{\bxi,m} = \left\{ \bxi \in \RR^m : \bW \bxi = \bt, \quad \bxi > 0 \right\},
\label{eq:constraint_xi}
\ee 
where $\bW \in \RR^{q \times m}$ has full row rank $q < m$, and $\bt \in \RR^q$. We assume $\bTheta_{\bxi}$ is non-empty and contains a strictly interior point $\bxi^\circ > 0$ such that
\[
\bW \bxi^\circ = \bt.
\]
This assumption can be checked as this is  a conic  membership problem  and a nonnegative solution $\bxi^\circ$  exists iff $\bt \in cone(\bW)$, the cone generated by the columns of $\bW.$ In the context of the SAE application, the matrix $\bW$ will be more like a weight matrix with nonnegative  entries and the external benchmark estimates $\bc$ will be also nonnegative. For a strictly positive solution in the interior of $\mcM_{\bxi,m}$,  the solution $\bxi^\circ$ has to be in the relative interior of $cone(\bW).$ Existence of $\bxi^\circ$ can be easily checked by linear programming. We now describe  how to  generate a bijection from $\mcM_{\bxi,m}$ to $\RR^{m-q}$ and using the bijection we invoke a prior $\pi_{\bxi}(\bxi)$ on $\mcM_{\bxi,m}$ and hence on $\mcM_m.$

\subsection{Exponential Paramterization Under Affine Constraints}
Let $r = m - q$ and let $\bB \in \RR^{m \times r}$ be a full column rank matrix with its column space equal to the null space of $\bW$, i.e., 
$\mcC(\bB) = \ker(\bW). $ To define the unconstrained parameter $\beeta \in \RR^r,$ we parameterize $\bxi$ as
\[
\bxi(\beeta) = \bxi^\circ \odot \exp\left( \bB \beeta - \bW^\top \blambda \right),
\]
where $\blambda := \lambda(\beeta) \in \RR^q$ is chosen such that
\be
\bW \bxi(\beeta) = \bt.
\label{eq:benchmark_eta}
\ee
Thus,
\[
\bxi_i(\beeta) = \bxi_i^\circ \exp\left( (\bB\beeta)_i - (\bW^\top \blambda)_i \right), \quad i=1,\dots,m.
\]
It is straightforward to see that the  exponential parameterization satisfies  assumptions (G1) and (G2).  The  following result establishes  (G3) and (G4). For any given  $\beeta$ there exists a unique  $\blambda$ which enforces \eqref{eq:benchmark_eta}. 
\begin{Proposition}
\lb{prop:unique-soln}
For each $\beeta \in \RR^r$, there exists a unique $\blambda = \lambda(\beeta)$ such that $\bW\bxi(\beeta) = \bt.$
\end{Proposition}
The optimization problem which is required to solve to obtain unique $\blambda,$ (see the  proof of \textbf{Proposition 4} in the appendix)  is  generally low dimensional since $q$ is usually small. The convex problem is strictly convex, making obtaining the solution extremely fast and the step can be embedded within a Bayesian sampling scheme. Let 
\be
h(\beeta,\blambda) = \bW\left( \bxi^\circ \odot \exp(\bB\beeta - \bW^\top \blambda) \right) - \bt
\lb{eq:h_fn}
\ee
We can  also use a simple Newton root solver to solve for $\blambda$ as the root of $h(\beeta,\blambda) = 0$ for a given $\beeta.$ Let at the $s$th iteration the variable be $\blambda^{(s)}.$ Then the iterations are
\beqa
\bxi^{(s)} &=& \bxi^\circ \odot \exp{(\bB\beeta - \bW^\top \blambda^{(s)})}, \\
h^{(s)}  &=& \bW\bxi^{(s)} - \bt, \\
H^{(s)} &=& \bW\diag(\bxi^{(s)})\bW^\top.
\eeqa

\section{Full Bayesian Inference}
Under the transformed parameters, the  model can be written as 
\[
\log \bY \mid \beeta \sim N\left( \log \bxi(\beeta), \bSigma_e \right),
\]
where $\beeta$ is the free parameter that describes the global parameterization in $\RR^{m-q}$.  Using the parameterization, we have $\log \bxi(\beeta) =
\log \bxi^\circ + \bB\beeta - \bW^\top \lambda(\beeta).
$ Then a hierarchical  prior  structure can be prescribed as 
\[
\beeta \mid \bbeta, \sigma_u^2 \sim N_r({\tilde{\bX}} \bbeta, \sigma_u^2 \bI_r),
\]
where ${\tilde{\bX}} = \bB^\top\bX$ and $\bX$ is a known design matrix. We further assume the following:
\[
\pi(\bbeta) \propto 1, \quad \sigma_u^2 \sim \mathrm{IG}(a_0, b_0).
\]
The posterior, of course, is not in closed form, but does admit partial conjugacy.  Specifically, the posterior is
\[
\pi(\beeta,\beta,\sigma_u^2 \mid \bY)
\propto
p(\log \bY \mid \beeta)\, p(\beeta \mid \bbeta,\sigma_u^2)\, \pi(\bbeta)\, \pi(\sigma_u^2).
\]
Explicitly,
\beqn
\pi(\beeta,\bbeta, \sigma_u^2 \mid \bY)
& \propto
& \exp\left(
-\frac{1}{2}(\log \bY - \log \bxi(\beeta))^\top \bSigma_e^{-1}(\log \bY - \log \bxi(\beeta))
\right) \nonumber\\
&& \times (\sigma_u^2)^{-(a_0 + r/2 + 1)}
\exp\left(
-\frac{1}{2\sigma_u^2} \|\beeta - {\tilde{\bX}} \bbeta\|^2 - \frac{b_0}{\sigma_u^2}
\right).
\label{eq:posterior}
\eeqn 
We will sample $\beeta$ from the posterior to make inferences on the quantities of interest $\bxi(\beeta)$ and $\log \bxi(\beeta)$. The benefit is that by construction the posterior of $\bxi(\beeta)$ based on \eqref{eq:posterior} is fully supported on $\mcM_{\bxi,m}$ and hence, the posterior of $\btheta$ is fully supported on the manifold $\mcM_m.$

\subsection{Posterior  Computation}
\label{PosteriorComputation}
This formulation is a  generalization of the standard logistic-normal  formulation on the simplex to the affine space given by $\bW\bxi = \bt.$
Analogously, we can use a  Metropolis within Gibbs sampler to sample from the posterior. Due to partial conjugacy some of the  full conditionals are explicitly given. 
\begin{enumerate}
\item Sample $\bbeta$ from its  full conditional.
\item Sample $\sigma_u^2$ from its full conditional.
\item Sample $\beeta$ using Metropolis--Hastings or gradient based sampler targeting $\pi(\beeta \mid  \bbeta,\sigma_u^2, \bY)$:
\begin{itemize}
\item For proposed $\beeta^*$, solve $\blambda(\beeta^\star)$ via Newton solving 
$h(\beeta^*, \blambda) = 0$  where $h$ is defined in \eqref{eq:h_fn}.
\item Compute $\bxi(\beeta^*)$ and $\log \bxi(\beeta^*)$.
\item Evaluate acceptance probability.
\end{itemize}
\end{enumerate}
The conditional for $\beeta$ needs to be sampled by an efficient sampler. Fortunately, the gradient for the target conditional can be explicitly evaluated, speeding up convergence. \\
\noindent{\bf Full Conditional Distributions:}
The full conditionals for Gibbs sampling are given as follows:
\begin{itemize}
\item {\bf Conditional for $\bbeta$:}
\[
\bbeta \mid \beeta, \sigma_u^2, \bY
\sim
N\left( ({\tilde{\bX}}^\top {\tilde{\bX}})^{-1} {\tilde{\bX}}^\top \beeta,\; \sigma_u^2 ({\tilde{\bX}}^\top {\tilde{\bX}})^{-1} \right).
\]
\item {\bf Conditional for $\sigma_u^2$:}
\[
\sigma_u^2 \mid \beeta, \bbeta, \bY
\sim
\mathrm{IG}\left(
a_0 + \frac{r}{2},\;
b_0 + \frac{1}{2} \|\beeta - {\tilde{\bX}} \bbeta\|^2
\right).
\]
\item {\bf Conditional for $\beeta$:}
\beqa 
\pi(\beeta \mid \bbeta,\sigma_u^2, \bY)
&\propto&
\exp\left(
-\frac{1}{2}(\log \bY - \log \bxi(\beeta))^\top \bSigma_e^{-1}(\log \bY - \log \bxi(\beeta))\right) \\
&& \times \exp\left(-\frac{1}{2\sigma_u^2} \|\beeta - {\tilde{\bX}} \bbeta\|^2
\right).
\eeqa
This distribution is non-Gaussian and must be sampled via Metropolis--Hastings or gradient-based methods. For efficient sampling the gradient can be evaluated explicitly. Define
$r(\beeta) = \log \bY - \log \bxi(\beeta).$ Then
\[
\nabla_{\beeta}  \log \pi(\beeta \mid \cdot)
=
\left( \frac{\partial \log \bxi}{\partial \beeta} \right)^\top \left[ \bSigma_e^{-1} r(\beeta)\right ]
- \frac{1}{\sigma_u^2} (\beeta - {\tilde{\bX}} \bbeta),
\]
To derive the gradient $\frac{\partial \log \bxi}{\partial \beeta}$ we can differentiate
$\bxi(\beeta) = \bxi^\circ \odot \exp(\bB\beeta - \bW^\top \blambda(\beeta)).$
Let $\bXi = \diag(\bxi(\beeta))$. 
Then differentiation gives
$
d\bxi = \bXi \left( \bB\, d\beeta - \bW^\top d\blambda \right). $
Since $\bW \bxi = \bt$, differentiating both sides we have $\bW\, d\bxi = \bzero,$
which implies $\bW \bXi (\bB\, d\beeta - \bW^\top d\blambda) = \bzero.$ Thus,
$\bW \bXi \bW^\top d\blambda = \bW \bXi \bB\, d\beeta,$
and hence
\[
\frac{\partial \blambda}{\partial \beeta}
=
(\bW \bXi \bW^\top)^{-1} \bW \bXi \bB.
\]
Substituting back, we have the gradient as 
\[
\frac{\partial \bxi}{\partial \beeta}
=
\bXi \left[ \bB - \bW^\top (\bW \bXi \bW^\top)^{-1} \bW \bXi \bB \right] = \bXi \left[ \bI - \bP_{\bW}\right]\bB
\]
where $\bP_{\bW} = \bW^\top (\bW \bXi \bW^\top)^{-1} \bW \bXi$ is the projection to $\mcC(\bW^\top)$  with respect to $\bXi.$
Thus, 
\[
\frac{\partial \log \bxi}{\partial \beeta}
= (\bI_m - \bP_{\bW})\bB.
\]
\end{itemize}

\subsection{Intrinsic Posterior Summary}

The proposed framework yields posterior draws of  $\btheta^{(s)}, s = 1,\ldots, S,$  of $\btheta$ that satisfy the benchmarking constraints, hence each $\btheta^{(r)}$, lies in the manifold $\mcM_m.$ 
Since $\mcM_m$ is nonconvex, the usual  posterior  point estimate, the arithmetic mean of the
draws $S^{-1}\sum_{s=1}^S \btheta^{(s)}$ need not satisfy the benchmark. A constraint-preserving summary
can instead be obtained through an intrinsic or extrinsic posterior mean. More detailed description along with pros and cons of  extrinsic  versus intrinsic  analysis in  shape restricted data and statistical models can be found in \cite{BhattacharyaPatrangenaru2003,  BhattacharyaPatrangenaru2005, BhattacharyaBhattacharya2012, Pennec2006,  DrydenMardia2016}. 

Under the Riemannian metric induced by the Euclidean embedding of $\mcM_m$, the empirical intrinsic mean is
\[
\widehat\btheta_I =
\underset{\ba \in \mcM_m} \argmin \frac{1}{S}\sum_{s=1}^S
d_{\mcM,m}\bigl(\ba,\btheta^{(s)}\bigr)^2,
\]
where $d_{\mcM,m}$ denotes geodesic distance on $\mcM_m$. Although this summary respects the manifold geometry, its computation generally requires repeated geodesic calculations within a nonlinear optimization, and the minimizer need not be unique.

The benchmarking constraints are linear in terms of the intermediate area level parameter $\bxi = \exp(\btheta)$  which lie  in the convex set $\mcM_{\bxi,m}.$ Hence, posterior aggregates such as ${\widehat{\bxi}} = S^{-1}\sum_{s=1}^S \bxi^{(s)},$ where $\bxi^{(s)} = \exp(\btheta^{(s)}),$ automatically satisfy the benchmarking constraints. 
Thus, an  extrinsic  posterior  summary for the model parameter is 
\[  \widehat\btheta_E = \log({\widehat{\bxi}}).\]
The corresponding area-level mean estimates are benchmarked as well. Since each posterior draw of $\btheta$ is already in the manifold, posterior credible intervals obtained from these draws are valid. This gives a simpler alternative as an extrinsic mean under the exponential embedding $\btheta \to \exp(\btheta)$. It minimizes the empirical posterior loss
$\frac{1}{S}\sum_{s=1}^S \left\|
\exp\bigl(\btheta^{(s)}\bigr)-\exp(\ba)\right\|^2,$
for $\ba\in\mcM_m.$
Thus, $\widehat\btheta_E$ provides a computationally straightforward benchmark-preserving summary corresponding to squared-error loss on the original mean scale. We adopt this summary for $\btheta$ throughout.

\section{Numerical Illustration}
We study the empirical performance of the proposed fully Bayesian estimator, computed under the constrained benchmarking prior (CBP),  through a Monte Carlo simulation study. The number of Monte Carlo replicates is $B=500.$ 
For comparison, we use the unconstrained hierarchical Bayesian log-normal model (UHB), which does not necessarily satisfy the benchmarking constraints. 

We consider $m = \{20,40,100\}$ areas. In each case, we generate $p = 5$ dimensional covariates $x_i \in \mathbb{R}^{p}$. We fix the set of covariates across the replications. The true area-level log means follow the linking model
\[
  \theta_i = x_i^\top \beta + u_i, \qquad u_i \stackrel{\text{iid}}{\sim} N(0,\sigma_u^2),
  \qquad \xi_i = \exp(\theta_i),
\]
with $\beta = {(0.5, 0.4, -0.3, 0.6, -0.2)^\top}$ and $\sigma_u^2 = 2$. Here $Y_i$ are direct survey estimates of the area totals, generated as 
\[
  \log Y_i = \theta_i + \epsilon_i, \qquad
  \epsilon_i \stackrel{\text{ind}}{\sim} N(0, \sigma_{i,e}^2),
\]
where the design  variances are generated once as $ \sigma_{i,e}^2 \sim \mathrm{Unif}[0.2,1]$ and kept fixed throughout the simulation. 
Following the standard practice in area-level models, sampling variances are treated as known and kept fixed across iterations.

For benchmarking, we consider two different configurations which lead to two different benchmark matrices. First, we consider a single benchmarking constraint on the sum of all the area-level means; hence $q=1,$  $A = (1,1,\cdots,1)$ and $t$ is a known scalar benchmark value. Second, we consider multiple non-overlapping benchmarking constraints, which mimic the case of benchmarking for regional totals with non-overlapping regions. More specifically, we assume $m_1, \cdots, m_q$ disjoint small areas involved with $\sum_{i=1}^q m_i =m$; hence $\bA = diag (1_{m_1}^\top, \cdots, 1_{m_q}^\top).$ For this exercise $m_1 = \cdots = m_q = \frac{m}{q};$ hence $A = I_q \otimes \mathbf{1}_{m/q}^\top.$ In either case, the matrix $A$ is full row rank.  Since we assume that the estimates at the aggregated level are observed with negligible sampling error, for simulation purposes we take $\bt$ to be the aggregation of true area-level totals. 
We consider two data-generating scenarios:
\begin{itemize}
  \item \textbf{Setting I (correct specification).} The fitted model uses all $p$
        covariates  that are involved in the data generating process.
  \item \textbf{Setting II (misspecified model: omitted covariate).} Covariates $x_4, x_5$ are
        generated but omitted from the fitted model.
\end{itemize}
For computation of the CBP posterior, we note from section~\ref{PosteriorComputation} that the hyperparameters ${\bbeta,\sigma_u^2}$ have closed-form conditionals, and for $\beeta$, the conditional distribution is non-Gaussian and hence gradient-based samplers might be employed for efficient sampling. Specifically, we use the Gibbs within Metropolis-Adjusted Langevin Algorithm (MALA) sampler where the Gibbs steps are utilized for ${\bbeta,\sigma_u^2}$ and the MALA step is used for updating $\eta$. At every proposal, $\lambda(\eta)$ is obtained by damped Newton iterations on the strictly convex objective as given in the proof of Proposition~\ref{prop:unique-soln}. During burn-in, the MALA step size is adapted on the log scale by Robbins--Monro updates toward an acceptance rate of $0.574$ \cite{roberts1998optimal}; the step-size is held fixed subsequently. Each chain runs
${50{,}000}$ iterations, with the first ${20{,}000}$ values discarded as burn-in.

To assess the performance of our proposed benchmarked estimator, we focus on the following metrics: the mean squared error (MSE), the coverage probability (CP), the average length (AL), and the interval score (IS) \cite{Gneiting01032007} of the credible intervals. MSE helps us assess the effect of imposing a benchmarking constraint inside the model, whereas CP, AL, and IS help assess the effect of benchmarking constraints on the calibration of credible intervals. We compute the MSE for both the underlying latent parameter level $\btheta$ and more importantly for the target parameters, i.e., the area-level mean values in the original scale. To calculate the MSE of the area-level means, we note that under the considered model, the true area mean for the $i-th$ area is given by $\mu_i = \exp(\theta_i + \frac{1}{2}\sigma_{e,i}^2)$, where $\theta_i$ is the true underlying area level parameter. Further, $\mu_i$ can be estimated by $\hat{\mu_i} = \hat{\xi_i}\exp{(\frac{1}{2}\sigma_{e,i}^2)} $, where  $\hat{\xi_i}$ denotes the posterior mean of $\xi_i$ under the given method. Hence the MSE at area mean level is defined as $\frac{1}{m} \sum_{i=1}^m (\mu_i - \hat{\mu_i})^2.$ Furthermore, for the unconstrained model, we compute the posterior summary of $\btheta$  in the same way as described in the section on intrinsic posterior summary. Specifically, we first obtain the posterior mean $\hat{\bxi}$ of $\bxi$, and then consider  \( \widehat\btheta = \log({\widehat{\bxi}}).\) as the posterior summary for $\btheta.$ Note that the intrinsic summary is not essential for the unconstrained model;  however, we adopt the same procedure to compute posterior summary for the constrained and the unconstrained models to ensure that both models are compared on the same scale.  

\begin{table}[H]
\centering
\caption{MSE for the small-area parameter ($\theta$) and for the area-mean ($\mu$) across Setup I (correctly specified) and Setup II
(omitted-variable misspecification) for the Direct estimator, unconstrained hierarchical
Bayes (UHB), and the proposed constrained benchmarking prior based estimator (CBP) with
$q = 1$ and $q = 5$ constraints. Results are based on
500 replications.} 
\label{tab:mse}
\begin{tabular}{llcccccc}
\toprule
 & & \multicolumn{3}{c}{Setup I} & \multicolumn{3}{c}{Setup II} \\
\cmidrule(lr){3-5} \cmidrule(lr){6-8}
Parameter & Model & $m=20$ & $m=40$ & $m=100$ & $m=20$ & $m=40$ & $m=100$ \\
\midrule
\multirow{4}{*}{$\theta$}
 & Direct        & 0.555 & 0.655 & 0.636 & 0.555 & 0.655 & 0.636 \\
 & UHB           & 0.542 & 0.619 & 0.570 & 0.638 & 0.712 & 0.676 \\
 & CBP ($q=1$)   & 0.518 & 0.641 & 0.589 & 0.496 & 0.634 & 0.585 \\
 & CBP ($q=5$)   & 0.380 & 0.569 & 0.559 & 0.368 & 0.572 & 0.559 \\
\midrule
% \multirow{4}{*}{$\xi$}
%  & Direct        & 12.66 & 13.85 & 3.47 & 12.66 & 13.85 & 3.47 \\
%  & UHB           & 15.06 & 6.06  & 1.54 & 14.91 & 13.73 & 2.30 \\
%  & CBP ($q=1$)   & 2.28  & 2.55  & 1.35 & 2.02  & 2.41  & 1.33 \\
%  & CBP ($q=5$)   & 1.09  & 0.46  & 0.93 & 1.00  & 0.49  & 0.92 \\
% \midrule
\multirow{4}{*}{$\mu$}
 & Direct        & 16.01 & 34.59 & 7.68 & 16.01 & 34.59 & 7.68 \\
 & UHB           & 19.10 & 14.83 & 3.36 & 19.03 & 34.10 & 5.02 \\
 & CBP ($q=1$)   & 3.12  & 5.93  & 2.94 & 2.77  & 5.55  & 2.89 \\
 & CBP ($q=5$)   & 1.53  & 0.90  & 2.02 & 1.41  & 0.96  & 1.99 \\
\bottomrule
\end{tabular}
\end{table}

\begin{table}[H]
\centering
\caption{Median MSE at the small-area parameter ($\theta$) level  and for the area-mean ($\mu$) level across Setup I and Setup II for the Direct estimator, unconstrained hierarchical
Bayes (UHB), and the proposed constrained benchmarking prior based estimator (CBP) with
$q = 1$ and $q = 5$ constraints. The MSE over areas is computed within each replication,
and the median is taken over the 500 replications.}
\label{tab:medmse}
\begin{tabular}{llcccccc}
\toprule
 & & \multicolumn{3}{c}{Setup I} & \multicolumn{3}{c}{Setup II} \\
\cmidrule(lr){3-5} \cmidrule(lr){6-8}
Parameter & Model & $m=20$ & $m=40$ & $m=100$ & $m=20$ & $m=40$ & $m=100$ \\
\midrule
\multirow{4}{*}{$\theta$}
 & Direct        & 0.529 & 0.646 & 0.629 & 0.529 & 0.646 & 0.629 \\
 & UHB           & 0.515 & 0.611 & 0.567 & 0.606 & 0.697 & 0.671 \\
 & CBP ($q=1$)   & 0.496 & 0.635 & 0.584 & 0.484 & 0.619 & 0.578 \\
 & CBP ($q=5$)   & 0.364 & 0.559 & 0.555 & 0.355 & 0.563 & 0.553 \\
\midrule
% \multirow{4}{*}{$\xi$}
%  & Direct        & 5.57 & 3.74 & 1.47 & 5.57 & 3.74 & 1.47 \\
%  & UHB           & 5.33 & 3.23 & 1.47 & 5.83 & 3.47 & 1.47 \\
%  & CBP ($q=1$)   & 1.11 & 2.06 & 1.28 & 0.95 & 1.83 & 1.24 \\
%  & CBP ($q=5$)   & 0.37 & 0.34 & 0.70 & 0.36 & 0.39 & 0.70 \\
% \midrule
\multirow{4}{*}{$\mu$}
 & Direct        & 7.12 & 9.03 & 3.22 & 7.12 & 9.03 & 3.22 \\
 & UHB           & 6.99 & 7.73 & 3.25 & 7.63 & 8.10 & 3.14 \\
 & CBP ($q=1$)   & 1.58 & 4.84 & 2.77 & 1.33 & 4.17 & 2.71 \\
 & CBP ($q=5$)   & 0.56 & 0.68 & 1.53 & 0.56 & 0.75 & 1.51 \\
\bottomrule
\end{tabular}
\end{table}
Table~\ref{tab:mse} and \ref{tab:medmse} report the mean and median MSE of the direct and model-based estimators across the latent parameter $\btheta$,  and the area mean $ \bmu$. The apparent non-monotonicty of the MSEs across $m$ is because the set of generated design variances, $d_i,$ had a bigger average for $m=40$ than the other two cases. Due to the highly skewed log-normal model, this made the MSEs of the estimators higher and thereby also providing a larger scope for improvement due to benchmarking. When the number of constraints is high, $q=5,$ and the number of areas is small, $m=20$, the effect of benchmarking is more pronounced for reduction of the MSE compared to estimators that are not benchmarked. When $m$ is higher, the  constrained and unconstrained estimators have similar performance with respect to $\btheta$. This is expected since, with a greater number of areas, a fixed set of benchmarking constraints carries comparatively less information about the true parameter values. With $q=1$, CBP shows an advantage over UHB for $m=20$ in the correctly specified model (Setup I), and for all values of $m$ in the misspecified model (Setup II). Under Setup II, the MSE of the UHB estimator is slightly lower than that of the CBP  with single constraints. Under the omitted-variable misspecification of Setup II, UHB deteriorates to the point of being worse than the direct estimator for all $m.$ However, the CBP estimators consistently outperform both the direct estimator and the UHB estimator. 
 One of the reasons for this is, of course, the misspecification and the fact that benchmarking constraints anchor the estimates to known true aggregates; however, no such additional information is provided for the UHB model. Another reason is the bias incurred due to the unconstrained summarization of the posterior for the  manifold valued  parameter. Recall that the reported estimate of $\btheta$ for the model-based estimators is
$\log \operatorname{E}(\bxi \mid \bY)$ and as a consequence, since the
logarithm is concave, by Jensen's inequality,
\(
  \log \operatorname{E}(\bxi \mid \bY) \;\geq\; \operatorname{E}(\log \bxi \mid \bY)
  \;=\; \operatorname{E}(\btheta \mid \bY),
\)
where the inequality and the logarithm are applied componentwise. Hence this
estimator is biased upward relative to the posterior mean of $\btheta$, and the
bias grows with the posterior dispersion of $\bxi$.  

The gains for CBP estimators are considerably larger for $\bmu$ scale. Exponentiation amplifies errors in areas with large values of $\theta$, and these are precisely the errors that the benchmark constraints by restricting the  $\xi$ parameter. Since the MSE for $\mu$ scales is computed after exponentiation, its mean over replications is sensitive to a small number of
replications with very large errors. We therefore also report the median over replications in Table~\ref{tab:medmse}. \par
\begin{figure}[H]
    \centering
    \includegraphics[width=0.95\linewidth]{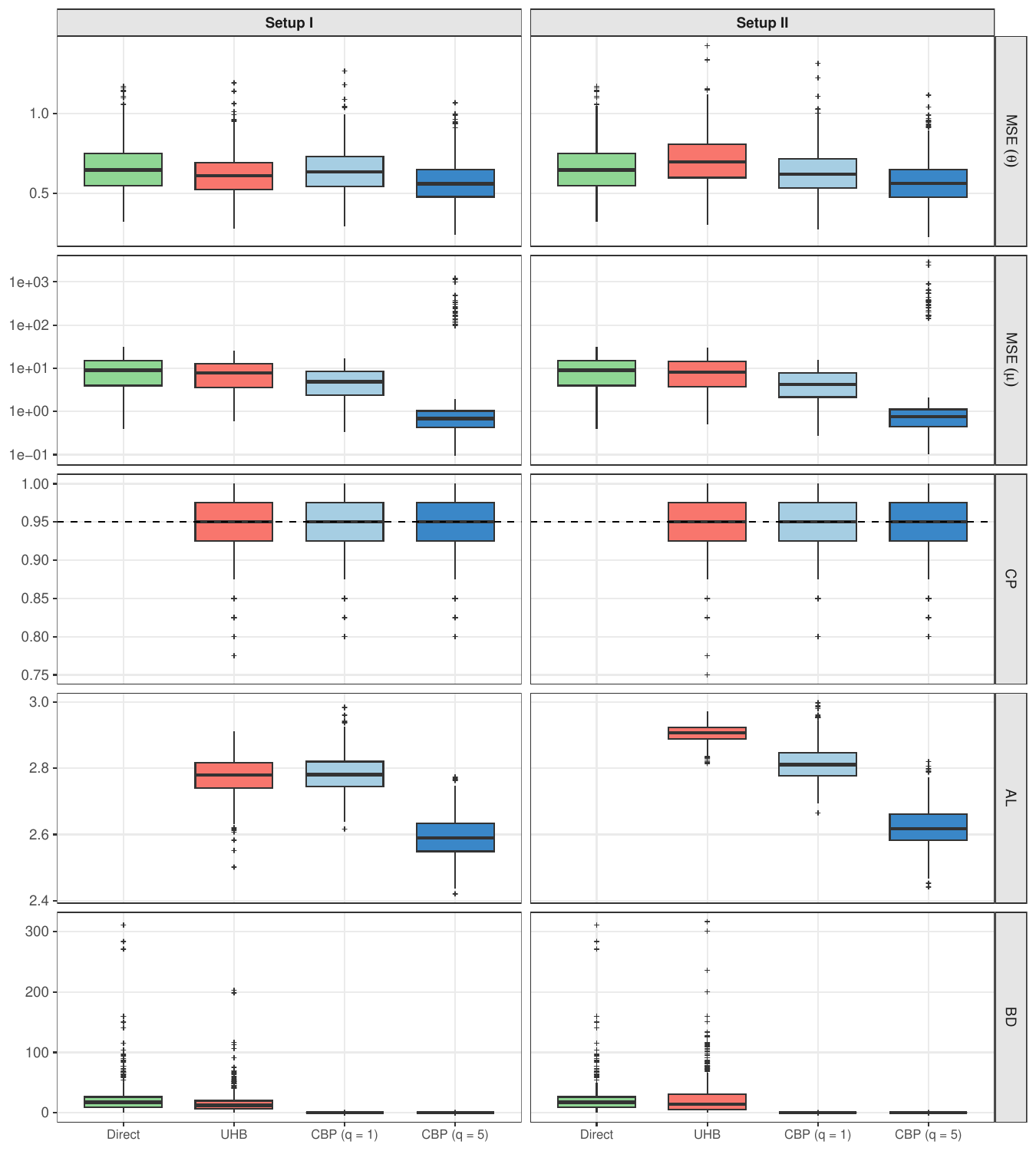}
    \caption{Distribution over 500 replications of the MSE at the $\theta$ and $\mu$ levels, the coverage probability (CP) and average length (AL) of 95\% intervals for $\theta$, and the benchmarking discrepancy (BD) for the Direct, UHB, CBP ($q=1$) and CBP ($q=5$) estimators with $m=40$,  under Setup~I and Setup~II. The dashed line marks the nominal 95\% coverage.} %\GSD{In the figure, the results for BD are stunningly dissimilar for non-benchmark and benchmark.} SB : THE BD/ MSE($\mu$) plots were in log scale. Updated with linear scale for BD
    \label{fig:boxplots}
\end{figure}

Table~\ref{tab:is} reports the interval score (IS) of the 95\% intervals, which rewards narrow intervals and penalizes intervals that fail to contain the true value. For $\theta$, all model-based intervals improve on the direct intervals in every setting, and CBP ($q=5$) attains the smallest IS throughout, with reductions relative to UHB of about 16\% in Setup~I and 19\% in Setup~II at $m=20$. As with the MSE, this advantage narrows as $m$ increases. 
\begin{table}[H]
\centering
\caption{Interval score (IS) of 95\% intervals at the small-area parameter ($\theta$) level across Setup I and Setup II
for the Direct estimator, unconstrained hierarchical
Bayes (UHB), and the proposed constrained benchmarking prior based estimator (CBP) with
$q = 1$ and $q = 5$ constraints. Results are based on
500 replications.}
\label{tab:is}
\begin{tabular}{llcccccc}
\toprule
 & & \multicolumn{3}{c}{Setup I} & \multicolumn{3}{c}{Setup II} \\
\cmidrule(lr){3-5} \cmidrule(lr){6-8}
Parameter & Model & $m=20$ & $m=40$ & $m=100$ & $m=20$ & $m=40$ & $m=100$ \\
\midrule
\multirow{4}{*}{$\theta$}
 & Direct        & 3.35 & 3.71 & 3.63 & 3.35 & 3.71 & 3.63 \\
 & UHB           & 3.14 & 3.36 & 3.27 & 3.19 & 3.49 & 3.39 \\
 & CBP ($q=1$)   & 3.07 & 3.35 & 3.27 & 3.01 & 3.34 & 3.28 \\
 & CBP ($q=5$)   & 2.65 & 3.12 & 3.18 & 2.59 & 3.13 & 3.20 \\
\midrule
% \multirow{3}{*}{$\mu$}
%  & UHB           & 7.34 & 7.48 & 4.22 & 7.69 & 8.20 & 4.33 \\
%  & CBP ($q=1$)   & 4.10 & 5.60 & 3.93 & 3.96 & 5.45 & 3.86 \\
%  & CBP ($q=5$)   & 2.61 & 2.80 & 2.87 & 2.56 & 2.90 & 2.87 \\
\bottomrule
\end{tabular}
\end{table}
Figure~\ref{fig:boxplots} displays the distribution over the 500 replications of the performance measures for $m=40$. The MSE panels confirm the observed patterns in Tables~\ref{tab:mse} and~\ref{tab:medmse}. At the $\theta$ level, the distributions are roughly symmetric with a
comparable spread across estimators.
CBP ($q=5$) is shifted downward relative to the others, and UHB shifts upward under the misspecification of Setup~II while the CBP distributions do not. At the $\mu$ level, displayed on the log scale, the direct and UHB estimators have long right tails, with a number of replications producing MSE one to two orders of magnitude above their medians. These replications are mainly responsible for the large mean MSE values in Table~\ref{tab:mse}. The CBP estimators have both lower medians and far lighter tails.

The coverage probability (CP) and average length (AL) panels shed more light on the interval score results of Table~\ref{tab:is}. All three model-based intervals
have median coverage close to the nominal 95\% level in both setups, with similar variability across replications. However,  CBP ($q=5$) produces significantly shorter intervals than UHB and CBP ($q=1$) in both setups without any loss of
coverage. This is the main reason for the lower interval score. Under
Setup~II, UHB maintains coverage only by widening its intervals, whereas the lengths of the CBP intervals are nearly unchanged from Setup~I. This can be attributed to the robustness of CBP to misspecification observed in the point
estimates. Finally, the benchmarking discrepancy (BD) panel shows that the direct and UHB estimators substantially violate the benchmark constraints in every replication. By contrast, the discrepancies for both CBP estimators satisfy the benchmarking constraints exactly.

\section{Real Data Analysis}
We illustrate the use of our model with an application motivated by the Small Area Income and Poverty Estimates (SAIPE) program using the American Community Survey (ACS) data produced by the U.S. Census Bureau \cite{USCensusSAIPE2023}. The ACS is a monthly survey that annually collects a sample of approximately 3.5 million addresses to produce estimates on a range of topics, some of which are not covered by the Decennial Census. Data are released for all counties in the U.S. in aggregated 5-year data products; however, the ACS only publishes county-level estimates for the 1-year data product for counties with total populations of at least 65,000 persons. For this article, we only focus on the county level child poverty estimates in the state of Maryland for illustration purposes. 

The SAIPE program produces model-based annual estimates of poverty and income for states, counties and school districts using estimates of poverty count data from the ACS. One outcome of interest is the count of children aged 5-17 in poverty, which represents school-aged children. The SAIPE program estimates rely on a small area model which includes covariates including SNAP participation (Supplemental Nutritional Assistance Program) and Federal Income Tax Returns \cite{BellEtAl2015}.

For this current application we introduce a synthetic, public use data set which is motivated by the Census Bureau's SAIPE production model at the county level. First, we utilize the publicly available 5-year ACS data product for 2023 as a stand-in for our 1-year survey estimates because these 5-year estimates are available for all counties in the U.S. This data will provide us with counts of children of age group $0-5$, $5-17$, and $0-17$ in poverty at the county level, as well as standard errors due to the survey. Next, we utilize the county-level SNAP count of participants for 2023 as a predictor in our regression model. The Federal Income Tax Return data is omitted as it is only available for internal Census Bureau use.

%{\color{blue}\textbf{Optional, can include if we want to look at scaled variances:}
We also incorporate additional modifications to our data to more closely mimic the 1-year ACS data used in the SAIPE production model. The 1-year ACS estimates have larger sampling errors compared to the 5-year ACS due to the smaller sample size in the 1-year product. The sample sizes for 1-year ACS data are approximately one-fifth of the sample sizes for the 5-year data; therefore, we can scale the variances from the 5-year product by 5 to get a "one-year" sampling variance with $\sigma_{i,1}^2 = 5*\sigma_{i,5}^2$. We then incorporate additive normal noise following $N(0, 4*\sigma_{i,5}^2)$ so that our synthetic one year count is $\tilde{Y}_{i,1} = Y_{i,5} + N(0, 4*\sigma_{i,5}^2)$.

For the context of benchmarking, we consider two sets of benchmark. First, within each county of Maryland the estimated poverty counts for the two age groups $0-5$ and $5-17$ years must sum to the five year ACS estimated poverty counts among children under 18. Next, the estimated counts at the county level for each age group must sum up to the estimated counts at the state level. Note that, with the age specific benchmarking constraints at the county level, considering aggregation for one age group from county to state level will be enough for this exercise. Specifically, we would put benchmarking constraints on the $0-5$ age group estimates, then estimates from $5-17$ age group will automatically benchmarked. 

 Let $y_{ij}$ denote the direct estimate of the poverty counts from $i$-th county and $j$-th age group, where $i = 1,\cdots, m$ and $j = 1,\cdots,l.$ For this particular application $m =24, l = 2.$ To fit the models jointly to two age groups, we stack the direct estimates by age groups, $\bY = (y_{11},y_{21},\cdots,y_{m1},$  $y_{12},y_{22},\cdots,y_{m2})^\top.$ The aggregation matrix $\bA$ is given by \(
  A =
  \begin{pmatrix}
    I_m & I_m \\[2pt]
    \mathbf{1}_m^\top & \mathbf{0}_m^\top
  \end{pmatrix}
  \in \mathbb{R}^{(m+1) \times 2m}
\). Furthermore, the aggregated total vector is denoted by $\bc \in \mathbb{R}^{m+1}$
so that the first $m$ elements give the county level counts under 18 from 5 year ACS and the last row gives the state level total for the age group $0-5$ years from 5 year ACS. The sampling variances on the log scale are obtained by the delta method. We set
\(
  \sigma^2_{ij} = \frac{s^2_{ij}}{y^2_{ij}},
\)
where $s_{ij}$ is the published standard error of
$y_{ij}$. Finally, since SNAP counts are available only at the county level, the values corresponding to each county are repeated for both age groups.   

% \begin{figure}[H]
%     \centering
%     % --- Left subfigure ---
%     \begin{subfigure}[t]{0.65\linewidth}
%         \centering
%         \includegraphics[width=\linewidth]{Plot_For_Count/md_estimatesv4.pdf}
%         \caption {Poverty Count Estimates}
%         \label{fig:estimate_map_plot}
%     \end{subfigure}
%     %\hfill
%     % --- Right subfigure ---
%     \begin{subfigure}[t]{0.765\linewidth}
%         \centering
%         \includegraphics[width=\linewidth]{Plot_For_Count/md_sev4.pdf}
%         \caption{Standard Error estimates}
%         \label{fig:se_map_plot}
%     \end{subfigure}
%     %
%     \caption{Statewise poverty count estimates and standard errors obtained from the CBP model}
%     \label{fig:benchmark_comparison}
% \end{figure}

\begin{figure}[htbp]
    \centering
        \includegraphics[width=0.75\linewidth]{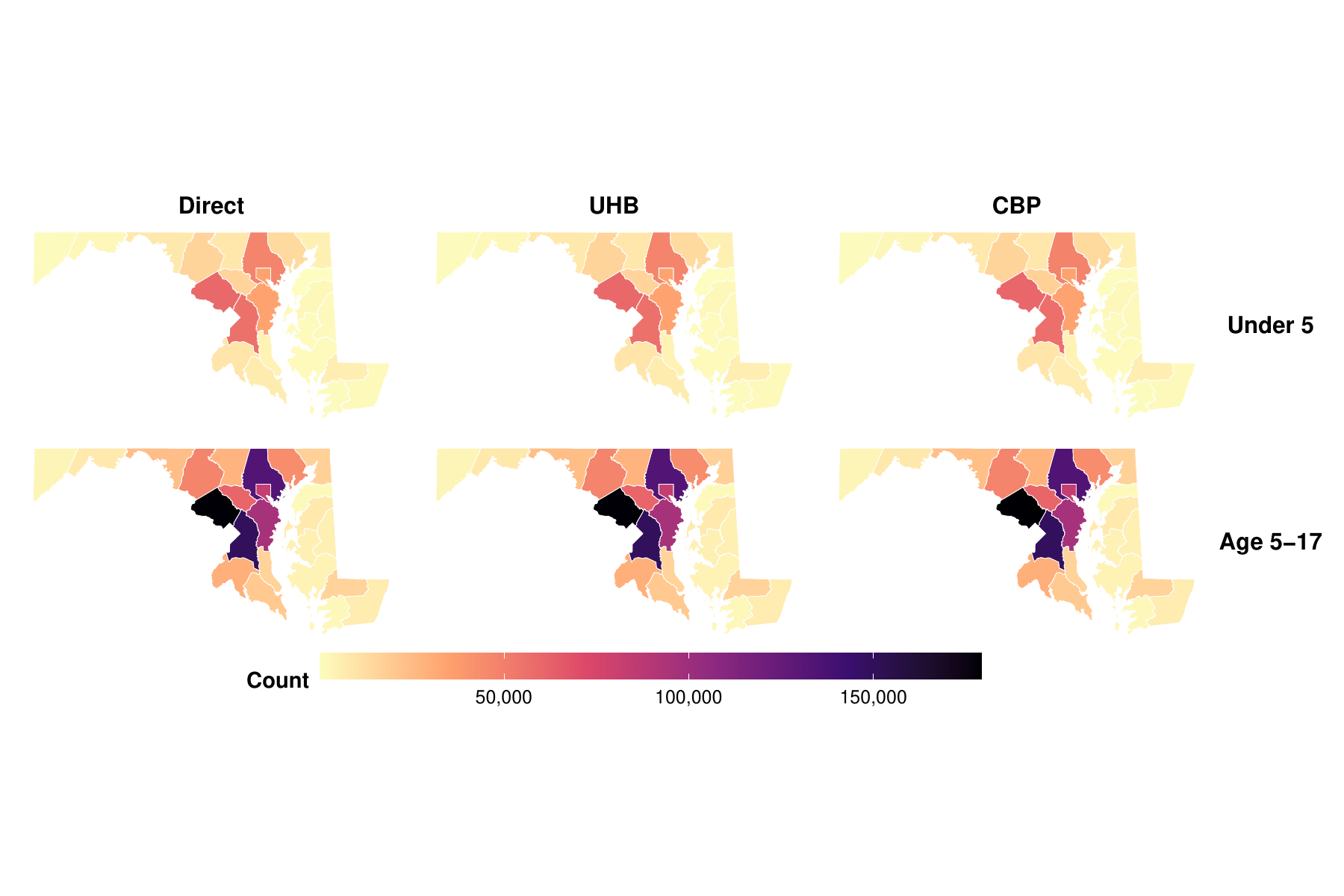}
        \caption {Poverty Count Estimates}
        \label{fig:estimate_map_plot}
\end{figure}

\begin{figure}[H]
    \centering
    % --- Left subfigure ---
    \begin{subfigure}[t]{0.40\linewidth}
        \centering
        \includegraphics[width=\linewidth]{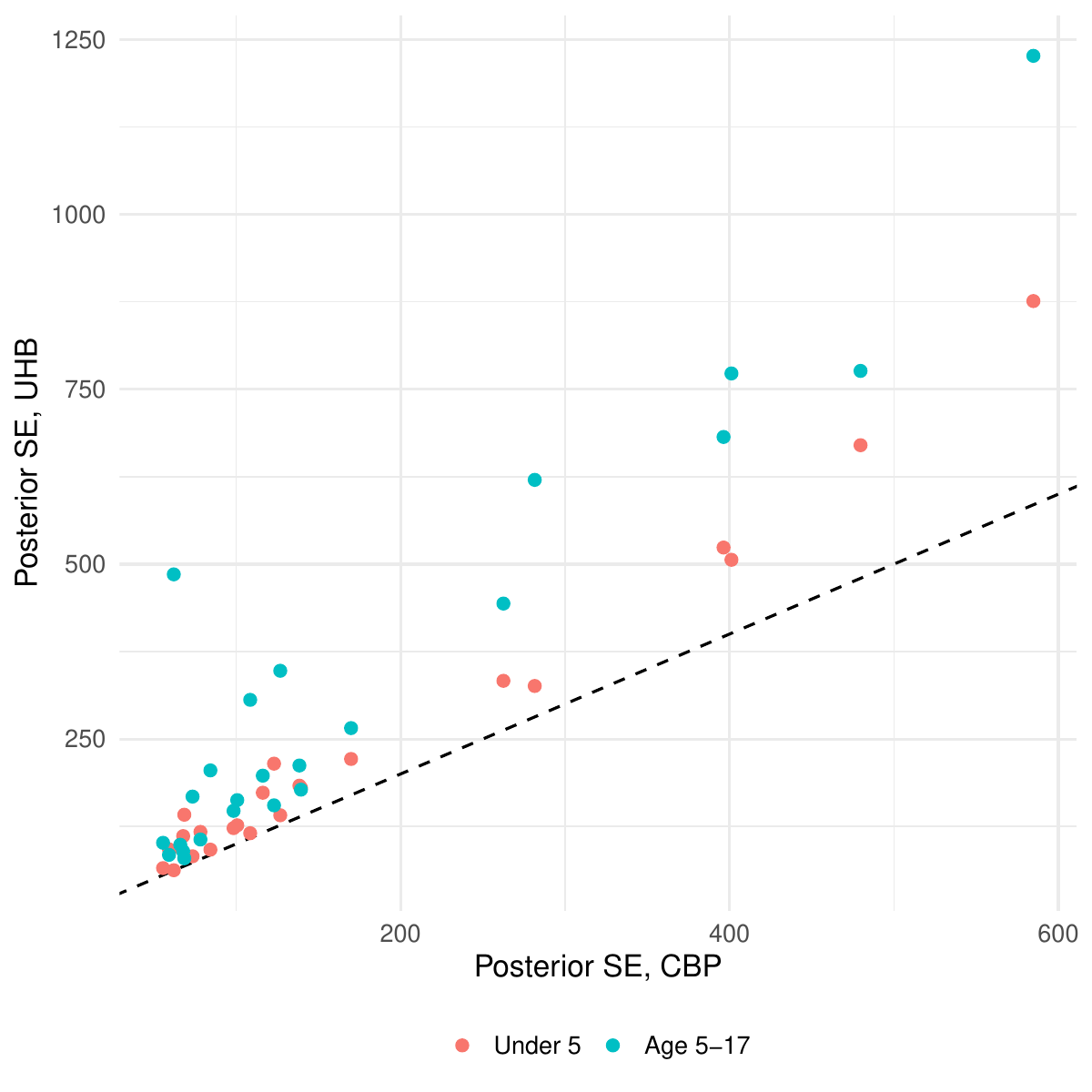}
        \caption {Comparison of standard errors}
        \label{fig:se_comp_plot}
    \end{subfigure}
    %\hfill
    % --- Right subfigure ---
    \begin{subfigure}[t]{0.40\linewidth}
        \centering
        \includegraphics[width=\linewidth]{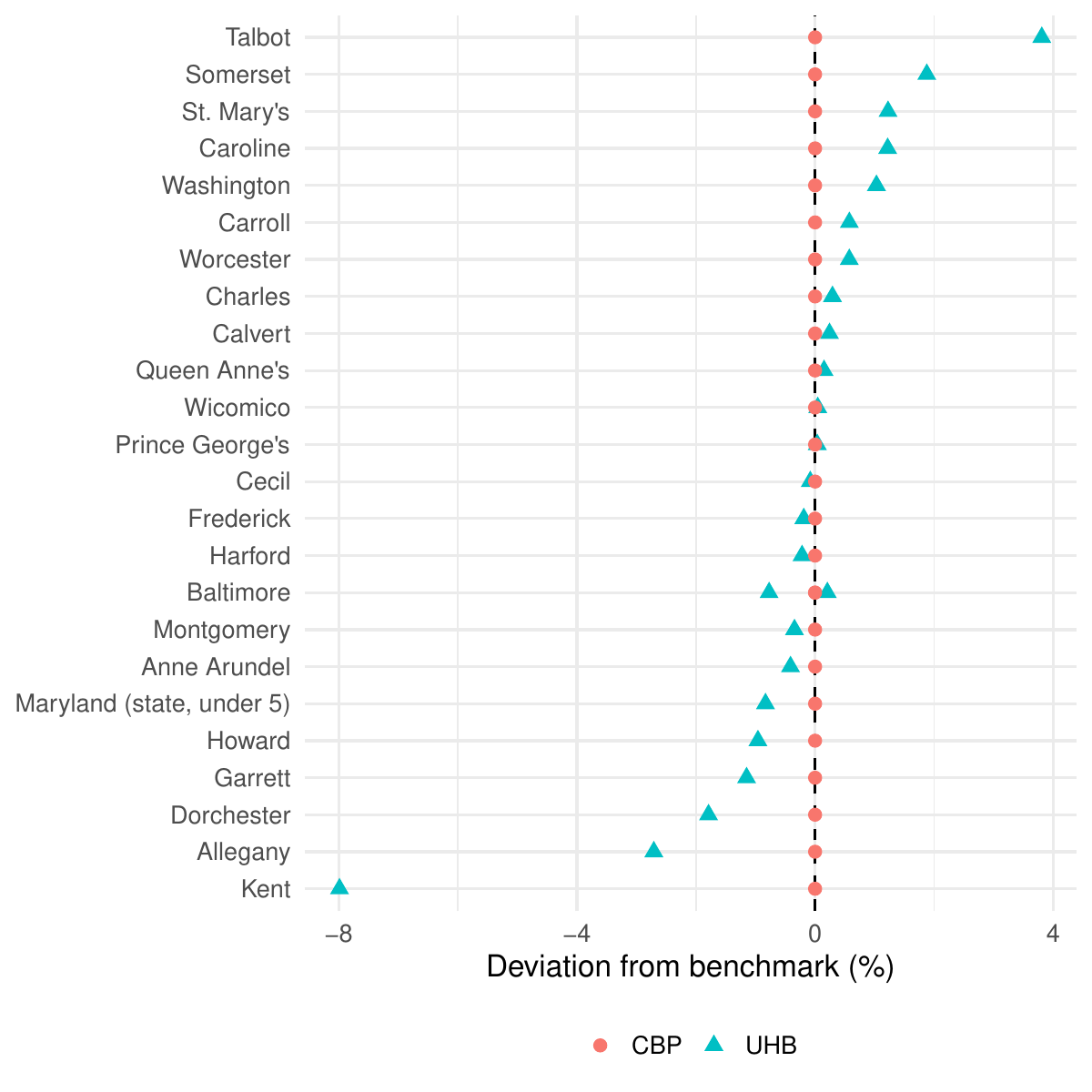}
        \caption{Benchmark Discrepancy}
        \label{fig:benchmark_discrepancy}
    \end{subfigure}
     \caption{ Comparison of UHB and CBP methods based on county-level posterior s.d.'s and benchmark discrepancies}
    \label{fig:benchmark_comparison}
\end{figure}

Figure~\ref{fig:estimate_map_plot} displays the county-level poverty count estimates for both age groups obtained from direct ACS estimates, the UHB model and the CBP model. Both model-based estimates show similar patterns as in direct estimates. The CBP and UHB model-based estimates are visually close in most counties. Figure~\ref{fig:se_comp_plot} compares the posterior standard errors of the UHB and the CBP estimates. Another interesting observation from Figure~\ref{fig:se_comp_plot} is that the CBP posterior SEs for the area-level mean $ \bmu$ for the $0-5$ age group and $5-17$ are the same. This follows from the benchmarking constraints since, for the $i$-th area  every posterior draw satisfies
$\mu_i^{< \text{ 5}} + \mu_i^{\text{5--17}} = t_i,$ where $t_i$  is equal to the known county-level aggregate, and hence corresponding standard errors should be equal to each other.
%\GSD{From the rightmost 14 points in Figure 3a it appears that the CBP posterior sd's for the <5 and 5-17 groups are the same. Is that right?}
For most counties, CBP estimates have lower standard errors, indicating that benchmarking improves the posterior uncertainty in the CBP estimates. Finally, Figure~\ref{fig:benchmark_discrepancy} reports, for each benchmark, the relative deviation of the aggregated count estimates from its target. As expected, UHB estimates do not satisfy the benchmarking constraints.The aggregated county totals for  children under 18 deviate from the five-year ACS totals by up to $8\%$, and the state total for children under $5$  deviated by $0.83\%.$ On the other hand, by construction, the CBP estimates satisfy every benchmark exactly, so that the county estimates for the two age groups add up to the published county totals and to the state
total.

\section{Discussion}
This article develops a fully Bayesian methodology for small area estimation under exact external benchmarking when the small area model is used to model a transformation of the area level direct estimates. Often these transformations facilitate Gaussian modeling, and hence the standard Gaussian Fay-Herriot model is used on the transformed scale. However, since the benchmarking restrictions are typically on the original scale, the constraints on the latent parameters become nonlinear. We accommodate these constraints by defining the prior directly on  smooth, lower-dimensional manifolds defined by the constraints. Consequently, the benchmark is incorporated into the hierarchical Bayesian model itself, and posterior draws automatically satisfy the external constraints, requiring no post hoc adjustment. Posterior inference therefore follows directly from Bayesian conditioning under the specified constrained model, providing a coherent framework for both point estimation and uncertainty quantification.
The simulation results demonstrate substantial reductions in mean squared error of the CBP estimates of area-level means relative to their unconstrained Fay-Herriot counterparts, with particularly pronounced gains when multiple  benchmarking constraints are imposed.
The application to synthetic data mimicking the SAIPE program illustrates the methodology for log-transformed count data. Through this application, we also illustrate how the proposed method can accommodate multi-dimensional or cross classified benchmarking constraints. The proposed method yields smaller posterior standard errors compared to the unconstrained FH model. These findings illustrate the potential efficiency gains from incorporating accurate external information directly into the hierarchical model. 

Although developed here for a logarithmic transformation with SAIPE as a motivating example, the framework applies more broadly. For example, logit or probit transformations can be used when the direct estimates are proportions subject to known population-weighted totals or averages. Log transformations can  accommodate positive rates, expenditures, or other
positive-valued area parameters whose weighted aggregates are externally specified. More generally, if  $\bpsi$ is a bijection from the original parameter space to a Euclidean space, external restrictions of the form
\[ \bW^\top \bpsi^{-1}(\btheta) = \bt \]
define a suitable constraint manifold whenever the corresponding constraint map is smooth and has a Jacobian of full row rank on the feasible set. This formulation also permits multiple, potentially overlapping, benchmarks provided the restrictions are compatible and satisfy the required regularity conditions.

% Future investigation will focus on large-sample behavior of the constrained posterior as the number of areas $m$ increases, including posterior concentration and the effect of estimating hyperparameters. Frequentist mean squared error comparisons with an appropriate oracle and with the unconstrained procedure will help characterize
% when the empirical efficiency gains extend to theoretical risk improvements. Impact of  the number and distribution of benchmark constraints on efficiency, the informativeness of the direct estimates, and the alignment of the external information with the underlying area-level structure will be investigated.

One limitation of the proposed methodology is that it treats the external benchmarks as exact. When these quantities are themselves estimated or subject to measurement error, extending the model to represent their uncertainty would provide a natural alternative to hard constraints. Further work on computation and sensitivity to the manifold-supported prior can broaden the practical applicability of the approach.

\section*{Appendix: Proofs of  Theorems and Propositions}

\appendix
\renewcommand{\theequation}{A.\arabic{equation}}
\setcounter{equation}{0}

\noindent{\bf Proof of Proposition~\ref{prop:regularity}}
\begin{proof}
Under (G2), $ \bzero \in \RR^q$ is a regular value of
$\bF$.  By the regular level set theorem, (see, e.g., \cite{Lee2013}) $\mcM_m = \bF^{-1}(\bzero)$ is locally a smooth submanifold of $\RR^m$ of dimension $\dim(\mcM_m) = m-q.$ 
The same theorem identifies the tangent space of a regular level set as the kernel of its derivative.  Hence,
\[
T_{\btheta}\mcM_m =
\ker D \bF(\btheta) =
\ker \left\{ \bW^\top J_{\bmu}(\btheta) \right\}.
\]
This proves the result.
\end{proof}

\noindent{\bf Proof of Proposition~\ref{prop:IFT-lambda}}
\begin{proof}
Differentiating \eqref{eq:Phi} with respect to
$\blambda^\top$ gives
\[
\frac{\partial\Phi(\beeta,\blambda)}
{\partial\blambda^\top} =
- \bW^\top J_{\bmu}(\btheta) \bC,
% \label{eq:Phi-lambda-Jac}
\]
where $\btheta = \btheta^{\circ} + \bB\beeta - \bC\blambda.$ At $(\beeta,\blambda) = (\bzero,\bzero)$, we have 
\[
\frac{\partial\Phi} {\partial\blambda^\top}
(\bzero,\bzero) = -\bW^\top J_{\bmu}(\btheta^{\circ})\bC.
\]
By (G1) and the  fact that $\msC(\bC) = \msC(\bW)$ we have $\det \left\{ \bW^\top J_{\bmu}(\btheta^{\circ}) \bC \right\} \neq 0.$
Since $\Phi(\bzero,\bzero)=\bzero$, the implicit function theorem
therefore yields neighborhoods $U$ and $V$ and a unique $C^1$ map $\blambda:U\rightarrow V$
satisfying
\[ \blambda(\bzero)=\bzero,  \mbox{ and } \Phi
\left\{ \beeta, \blambda(\beeta) \right\} = \bzero.
\]
Substitution into the definition of $\Phi$ gives
$\bW^\top \bmu \left\{ \btheta(\beeta)\right\}
= \bt,$ and  so $\btheta(\beeta)
\in\mcM_m$ for every $\beeta\in U$. \\
  
\end{proof}

\noindent{\bf Proof of Proposition~\ref{prop:H-tangent}}
\begin{proof}
Differentiating  \eqref{eq:chart} with respect to $\beeta^\top$ gives $\frac{\partial \btheta(\beeta)} {\partial\beeta^\top} = H(\beeta) = \bB - \bC \frac{\partial\blambda(\beeta)} {\partial\beeta^\top}.$ On the other hand, $F\{\btheta(\beeta)\} = \bzero$ for all $\beeta\in U$.  Differentiating this identity with respect to $\beeta^\top$ yields, by the chain rule,
\[
D \bF\{\btheta(\beeta)\} \frac{\partial \btheta(\beeta)} {\partial\beeta^\top} = \bzero.
\]
Thus $D \bF\{\btheta(\beeta)\} \bH(\beeta) = \bzero.$
Every column of $\bH(\beeta)$ therefore belongs to
$\ker D \bF\{\btheta(\beeta)\}.$ By Proposition \ref{prop:regularity}, $\ker D \bF\{\btheta(\beeta)\}
= T_{\btheta(\beeta)}\mcM_m.$ Hence, $\msC\{H(\beeta)\} \subseteq T_{\btheta(\beeta)}\mcM_m$. It remains to prove that $\bH(\beeta)$ has full column rank. Suppose
$\bH(\beeta)\ba =\bzero$  for some $\ba\in\RR^{m-q}.
$ Then $\bB\ba =  \bC \frac{\partial\blambda(\beeta)}
{\partial\beeta^\top} \ba.$ The left-hand side belongs to $\msC(\bB)$ and the right-hand
side belongs to $\msC(\bC)$.  Assumption (P1) implies
$\msC(\bB) \cap \msC(\bC) = \{\bzero\}.$
Therefore $\bB\ba=\bzero.$ Since $\bB$ has full column rank, $\ba = \bzero.$ Thus, the columns of $H(\beeta)$ are linearly independent, and hence,
\(
\operatorname{rank}\bH(\beeta)=m-q.
\)
By Proposition \ref{prop:regularity}, $\dim
T_{\btheta(\beeta)}\mcM_m = m-q.$ Since
$\msC\{\bH(\beeta)\} \subseteq T_{\btheta(\beeta)}\mcM_m$
and both spaces have dimension $m-q$, they are equal.  Hence,
\[
\mathcal C\{H(\beeta)\}
=
T_{\btheta(\beeta)}\mcM_m.
\]
\end{proof}

\noindent{\bf Proof of Theorem~\ref{thm:local-chart}}
\begin{proof}
By Proposition \ref{prop:IFT-lambda}, 
$\Theta$ is a $C^1$ map from a neighborhood of
$\bzero\in\RR^{m-q}$ into $\mcM_m$ and $\Theta(\bzero) =
\btheta^{\circ}.$ By Proposition \ref{prop:H-tangent},
$D\Theta(\beeta) = H(\beeta)$ has rank $m-q$, and
$\mathcal C\{H(\beeta)\} = T_{\Theta(\beeta)}\mcM_m.$ Thus
\[
D\Theta(\beeta)
:
\RR^{m-q}
\rightarrow
T_{\Theta(\beeta)}\mcM_m
\]
is one-to-one and onto.  Since both vector spaces have dimension $m-q$,
$D\Theta(\beeta)$ is a linear isomorphism. Now regard $\Theta$ as a smooth map between two manifolds of the same
dimension: $\Theta: \RR^{m-q} \to \mcM_m.$ At $\beeta=\bzero$, its derivative $D\Theta(\bzero) : \RR^{m-q}
\to T_{\btheta^{\circ}}\mcM_m$  is an isomorphism.  The inverse function theorem for manifolds therefore
implies that there exist neighborhoods $U_0\ni\bzero $
and $\mcV_0\ni\btheta^{\circ}$ such that $\Theta:U_0\rightarrow\mathcal V_0$ is a diffeomorphism.
Hence every point of $\mcM_m$ sufficiently close to
$\btheta^{\circ}$ is represented by exactly one nearby
$\beeta$, and the inverse coordinate map
\[
\Theta^{-1}:\mathcal V_0\rightarrow U_0
\]
is continuously differentiable.  Thus, $\beeta$ constitutes a
regular local coordinate system on $\mcM_m$.
\end{proof}

\noindent{\bf Proof of Theorem~\ref{thm:global-chart}}
\begin{proof}
We prove the result in several steps.

\paragraph{Step 1: smoothness.}

By assumptions (G3), for every $\beeta\in\RR^{m-q}$
there exists a unique $\blambda(\beeta)$  such that
$\Phi \left\{ \beeta, \blambda(\beeta) \right\} = \bzero.$
From the proof of Proposition~\ref{prop:IFT-lambda}, the derivative of $\Phi$ with respect to $\blambda^\top$ is
\[ \frac{\partial\Phi} {\partial\blambda^\top} =
-\bW^\top J_{\bmu}(\btheta) \bC, \]
where $\btheta = \btheta^{\circ} + \bB\beeta -\bC\blambda.$
By (G4), this matrix is nonsingular at every solution.  Hence, by the implicit function theorem, around every
$\beeta_0$ there exists a neighborhood on which the solution
$\blambda(\beeta)$ is $C^1$. Because the solution is globally unique, these local solution maps agree on all overlaps.  They therefore patch together to produce a globally defined $C^1$ map
$\blambda: \RR^{m-q} \to \RR^q.$

\paragraph{Step 2: injectivity.}

By construction,
$\Phi\left\{\beeta,\blambda(\beeta)\right\}=\bzero.$ Therefore $\bW^\top\bmu \left\{ \Theta(\beeta)\right\}=
\mathbf t.$ Hence $\Theta(\beeta) \in\mcM_m$ for every
$\beeta\in\RR^{m-q}.$ Suppose $\Theta(\beeta_1) = \Theta(\beeta_2).$ Then
\[
\bB(\beeta_1-\beeta_2) = \bC \left[ \blambda(\beeta_1) -
\blambda(\beeta_2) \right].
\]
Thus, the same vector belongs simultaneously to $\msC (\bB)$
and  $\msC(\bC).$ Since $[\bB\;\bC]$ is nonsingular,
$\msC(\bB) \cap \msC(\bC) = \{\bzero\}. $ Therefore
$\bB(\beeta_1-\beeta_2) = \bzero.$ Since $\bB$ has full column rank, $\beeta_1 = \beeta_2.$ Hence $\Theta$ is injective.

\paragraph{Step 4: surjectivity.}

Let $\btheta\in\mcM_m.$ Since $[\bB\;\bC]$ is nonsingular, there exist unique $\beeta\in\RR^{m-q},$ and $\blambda\in\RR^q$ such that
\be
\btheta - \btheta^{\circ} = \bB\beeta - \bC\blambda.
\lb{eq:general-decomp}
\ee
Because $\btheta\in\mcM_m,$ we have $\bW^\top
\bmu(\btheta) = \bt.$ Substitution of \eqref{eq:general-decomp} gives $\Phi(\beeta,\blambda) = \bzero. $ By the global uniqueness assumption in (G3), $\blambda = \blambda(\beeta).$ Therefore $\btheta = \Theta(\beeta).$
Hence $\Theta$ is surjective.

\paragraph{Step 5: smoothness of the inverse.}

Since $[\bB\;\bC]$ is nonsingular, define $[\bL_B;\bL_C]^\top = [\bB\;\bC]^{-1},$ partitioned so that $\bL_B^\top\bB=I_{m-q},$ and $\bL_B^\top\bC=0.$ Then from
\[ \btheta - \btheta^{\circ} = \bB\beeta - \bC\blambda(\beeta), \]
multiplication by $\bL_B^\top$ gives $\beeta = \bL_B^\top
(\btheta-\btheta^{\circ}).$ Thus
\[
\Theta^{-1}(\btheta) = \bL_B^\top (\btheta-\btheta^{\circ}),
\]
which is smooth. Therefore, $\Theta$ is a global $C^1$ diffeomorphism from
$\RR^{m-q}$ onto $\mcM_m$.

By construction, $[\bB\;\bC]$ is nonsingular. Thus every
$\bv \in \RR^m$ admits a unique decomposition $\bv = \bB\beeta - \bC\blambda.$ Hence, every $\btheta \in \mcM_m$ has a  unique  decomposition and the unrestricted variable $\beeta$ describes a global coordinate system. 
\end{proof}

\noindent{\bf Proof of Theorem~\ref{thm:posterior-equivalence}}
\begin{proof}
The Coarea formula applies directly to the vector-valued map $\bF : \RR^m \to \RR^q.$ 
By the Coarea formula, the resulting prior has density with respect to $(m-q)$-dimensional Hausdorff measure on $\mcM_m$ proportional to
\be
\pi_B(\btheta \mid \bbeta, \sigma_u^2) \propto
\frac{\pi_U(\btheta \mid \bbeta, \sigma_u^2)I(\btheta\in\mcM_m)}{
J_F(\btheta)}
\label{eq:conditional-prior}
\ee
where $J_F(\btheta)$ is the Coarea Jacobian 
$J_F(\btheta) =  \left\{\det\left[D\bF(\btheta) D\bF(\btheta)^\top\right]\right\}^{1/2}.$ 
Differentiating the  benchmarking equations, we get 
\[
D\bF(\btheta) = \bW^\top J_{\mu}(\btheta),
\]
Since $J_{\mu}(\btheta) =  \diag\left\{\mu_1'(\theta_1),\ldots,\mu_m'(\theta_m)^2\right\}$,  we have 
\[
J_F(\btheta) = \left\{\det\left[\bW^\top \diag\left\{\mu_1'(\theta_1)^2,\ldots,\mu_m'(\theta_m)^2
\right\}\bW\right]\right\}^{1/2}.\]
Because $\bH(\beeta) = D\btheta(\beeta)$ has full column rank $m-q$, the Hausdorff surface element induced by the chart is
\[ d\mcH_{m-q}\{\btheta(\beeta)\} = J_{\theta}(\beeta)\,d\beeta, \]
where $J_{\theta}(\beeta) =
\left\{ \det \left[\bH(\beeta)^\top \bH(\beeta)\right]\right\}^{1/2}$ is the Jacobian.
Combining this with \eqref{eq:conditional-prior} gives the coordinate density under the reduced coordinates as 
\be
\pi_{B,\beeta}(\beeta \mid \bbeta, \sigma_u^2)
\propto
\pi_U(\btheta(\beeta) \mid  \bbeta, \sigma_u^2))
\frac{J_{\theta}(\beeta)}{J_F\{\btheta(\beeta)\}}.
\label{eq:general-coordinate-prior}
\ee
Explicitly,
\[
\pi_{B,\beeta}(\beeta \mid \bbeta, \sigma_u^2)
\propto
\pi_U(\btheta(\beeta) \mid  \bbeta, \sigma_u^2)
\,
\frac{\left\{ \det \left[\bH(\beeta)^\top \bH(\beeta)\right]\right\}^{1/2}}{
\left\{\det \left[ \bW^\top J_{\mu}(\btheta(\beta))
J_{\mu}(\btheta(\beta))^\top \bW \right]\right\}^{1/2}}.
\]
The reduced-coordinate posterior is
\[
\pi_{B,\beeta}(\beeta \mid \bY,  \bbeta, \sigma_u^2)
\propto
\mcL(\bY \mid \btheta(\beeta))
\pi_{B,\beeta}(\beeta \mid \bbeta, \sigma_u^2).
\]
Using \eqref{eq:general-coordinate-prior} we have,
\beqa
\pi_B(\theta\mid \bY,\bbeta,\sigma_u^2)
&\propto & \mcL(\bY \mid \btheta(\beeta))
\pi_U(\btheta(\beeta) \mid  \bbeta, \sigma_u^2))
\frac{J_{\theta}(\beeta)}{J_F\{\btheta(\beeta)\}}
\nonumber\\
&\propto&
\pi_U(\btheta(\beeta) \mid \bY, \bbeta, \sigma_u^2))
\frac{J_{\theta}(\beeta)}{J_F\{\btheta(\beeta)\}}.
\label{eq:general-proof-coordinate}
\eeqa
Since $d\mcH_{m-q}\{\btheta(\beeta)\} = J_{\theta}(\beeta)\,d\beeta $ the preceding density is exactly the coordinate representation of
\[
\frac{\pi_U(\btheta\mid \bY,\bbeta, \sigma_u^2)
}{J_{F}(\btheta)}\,d\mcH_{m-q}(\btheta)
\]
on $\mcM_m$. By the Coarea formula, this is the regular conditional posterior given $\bF(\btheta) = \bzero.$ 
\end{proof}

\noindent{\bf Proof of Proposition~\ref{prop:unique-soln}}
\begin{proof}
Define
$
h(\beeta,\blambda) = \bW\left( \bxi^\circ \odot \exp(\bB\beeta - \bW^\top \blambda) \right) - \bt.
$
Define the convex function
\[
\Upsilon_{\beeta}(\blambda)
=
\bone^\top \left( \bxi^\circ \odot \exp(\bB\beeta - \bW^\top \blambda) \right)
+ \blambda^\top \bt.
\]
Then $\nabla_{\blambda} \Upsilon_{\beeta}(\blambda)
=
- \bW \left( \bxi^\circ \odot \exp(\bB\beeta - \bW^\top \blambda) \right) + \bt
= - h(\beeta,\blambda).
$
Hence we have 
\[  
\blambda  = \underset{\blambda \in \RR^q}\arg\min \;\Upsilon_{\beeta}(\blambda)
\label{eq:blambda}
\]
The Hessian is
\[
\nabla^2_{\blambda} \Upsilon_{\beeta}(\blambda)
=
\bW \, \diag(\bxi(\beeta)) \, \bW^\top,
\]
which is positive definite since $\bW$ has full row rank and $\bxi(\beeta) > 0$. Hence $\Upsilon_{\beeta}$ is strictly convex and admits a unique minimizer, which solves $h(\beeta,\blambda)= \bzero$.
\end{proof}

\bibliographystyle{plainnat}
\bibliography{references_count}

\end{document}